\documentclass[prb,amsmath,amssymb,superscriptaddress,twocolumn]{revtex4}
\usepackage{amsmath}
\usepackage{amsfonts}
\usepackage{amssymb}
\usepackage{mathrsfs}
\usepackage{bm}
\usepackage{dsfont}
\usepackage{array}
\usepackage{graphicx}

\newcommand{\angstrom}{\textup{\AA}}

\begin{document}

\title{Berry phases and the intrinsic thermal Hall effect in high temperature cuprate superconductors}

\author{Vladimir Cvetkovic}
\affiliation{National High Magnetic Field Laboratory, Tallahassee, FL 32306, USA}
\author{Oskar Vafek}
\affiliation{National High Magnetic Field Laboratory, Tallahassee, FL 32306, USA}
\affiliation{Department of Physics, Florida State University, Tallahassee, FL 32306, USA}
\email{vafek@magnet.fsu.edu}

\begin{abstract}
 The Bogoliubov quasiparticles move in a practically uniform magnetic field in the vortex state of high temperature cuprate superconductors. Do the quasiparticles experience a Lorentz force when set in motion by an externally applied heat current ${\bf j}_Q$, bending their trajectories and causing the temperature gradient perpendicular to ${\bf j}_Q$ and the applied field ${\bf H}$, or is the thermal Hall effect a consequence of Berry phases as in an intrinsic anomalous Hall effect of a semiconductor/metal with spin-orbit coupling? Here we show that it is the latter, and for the first time, calculate the temperature, ${\bf H}$-field and the $d$-wave pairing gap $\Delta$ dependence of the intrinsic thermal Hall conductivity, $\kappa_{xy}$. We find that the intrinsic contribution to $\kappa_{xy}$ displays a rapid onset with increasing temperature, which compares favourably with existing experiments at high ${\bf H}$-fields on the highest purity samples. This finding may help to settle a much-debated question of the bulk value of the pairing strength in cuprate superconductors in magnetic field.
\end{abstract}

\maketitle

The low energy excitations of a superconductor, the so called Bogoliubov quasiparticles (qps), are a quantum mechanically coherent superposition of an electron and a hole. As such, they do not carry a fixed electrical charge. In the vortex state of an extreme type-II superconductor, such as high T$_c$ cuprates, the externally applied magnetic field ${\bf H}$ penetrates the sample in the form of flux tubes, whose width, set by the penetration depth $\lambda$, is much larger than the intervortex separation set by the magnetic length $\ell_H=\sqrt{hc/(eH)}$, for essentially the entire range of ${\bf H}$-field applied in a typical experiment. Therefore, the qps move in a practically uniform magnetic (induction) field ${\bf B}$.
Among the best established physical properties of cuprate superconductors in the vortex state is the $\sqrt{H}$-increase of the low temperature ($T$) specific heat when the $H$-field is applied perpendicular to the CuO$_2$ planes. This effect was predicted by Volovik\cite{VolovikJETP1993} based on a semiclassical analysis of the qp motion in the vortex state of d$_{x^2-y^2}$ superconductor. Although such analysis successfully captures the overall $\sqrt{H}$ increase of the qp density of states, a full quantum mechanical solution is necessary in order to investigate non-trivial topological structure of the qp wavefunctions in the vortex state. Here we provide such solution. Using the Berry phases associated with the qp wavefunctions, we calculate, for the first time, the full $T$, ${\bf H}$-field and the $d$-wave pairing gap $\Delta$ dependence of the intrinsic contribution to the thermal Hall conductivity, thereby shedding light on physics beyond the Volovik effect.

In the thermal Hall effect a small applied energy current ${\bf j}_E$ causes a transverse temperature gradient $\nabla T$, i.e., a non-zero off-diagonal component of the Fourier law ${j_{E}}_{\mu}=-\kappa_{\mu\nu}\nabla_{\nu}T$. This effect has been reported experimentally in (slightly overdoped) YBa$_2$Cu$_3$O$_{6.99}$ samples by Ong and collaborators\cite{OngPRL2001}, and by Zeini {\it et.al.}\cite{Zeini1999}. The measurements extend up to $15$Tesla, and down to the lowest temperature of $\sim 10$Kelvin. At the largest $H$-fields, $\kappa_{xy}$ is positive, decreases with decreasing $T$. However, linear extrapolation to $T\rightarrow 0$ would yield finite {\em negative} value of $\kappa_{xy}$, which is unphysical because $\kappa_{xy}$ must vanish in such limit. In fact, the data at the highest fields is consistent with $\kappa_{xy}$ being very small bellow $T_{onset}\sim 7K$, followed by a rapid increase at $T_{onset}$ with an almost linear $T$-dependence at higher $T$. New (unpublished) data at lower doping and lower temperature are consistent with such onset.

Initial theoretical arguments of Simon and Lee\cite{SimonLee1997} were based on a continuum model, linearized in the vicinity of the Dirac points. They used Kubo formula to argue for the scaling form $\kappa_{xy}\sim T^2 F\left(k_BT\ell_H/\hbar v \right)$ where $v=\sqrt{v_F \Delta/p_F}$, $p_F$ and $v_F$ being Fermi momentum and velocity respectively and $\Delta$ being the $d$-wave gap parameter. Based on the same linearized continuum model, Vishwanath\cite{VishwanathPRL2001} argued that the qp spectrum in the vortex state acquires a small gap which grows linearly with ${\bf H}$, and that $\lim_{T\rightarrow 0}\kappa_{xy}/T=n\pi k^2_B/(6\hbar)$ with $n=0$ or $\pm 2$. However, the linearized approximation has been found to be problematic. The qp spectrum is not invariant\cite{OV2001a,MelikyanTesanovic2007} under the large (singular) gauge transformations\cite{FranzTesanovic2000,Marinelli2000}, which were being employed in the calculations\cite{VishwanathPRL2001} of $\kappa_{xy}$. Because such large gauge invariance must be present, the validity of the results has been in doubt. A tight-binding lattice regularization was found to remove such problems\cite{OV2001a}, and the initial model calculations\cite{OV2001b} of $\kappa_{xy}$ for unrealistically small values of $\ell_H$, corresponding to over $1700$ Tesla, indeed found that for a space inversion symmetric vortex lattice, the qp spectrum is generally gapped, and that $\lim_{T\rightarrow 0}\kappa_{xy}/T=n\pi k^2_B/(6\hbar)$, albeit the values of $n$ ranged from $-2$ to $12$. Later, index theoretic arguments\cite{OVAMprl2006} very near half-filling found commensuration based oscillations between $n=\pm 4$ and $n=0$.
Finally, using the linearized model and a number of drastic simplifying assumptions, the weak field $\kappa_{xy}\sim T\sqrt{H}$ was argued be due to thermally excited nodal qps, scattering primarily from impurities, with a small skew component due to vortices\cite{DurstPRL2003}.

Here we build on the observation\cite{OV2001b} that the intrinsic contribution to $\kappa_{xy}$ is a direct consequence of the qp wavefunctions acquiring a non-trivial Berry curvature\cite{Berry1984} upon adiabatic changes of the vortex crystal momentum, ${\bf k}$.
Because ${\bf j}_E$ is perpendicular to the thermal Hall gradient, the (non-dissipative) effect described here does not contribute to the entropy production, $dS/dt=-\int d^2{\bf r} (\nabla T/T^2)\cdot {\bf j}_E({\bf r})$. The effect can therefore be addressed without considering irreversible processes.
Without making any further simplifying assumptions, apart from treating the qps as non-interacting and the vortices forming a perfect (stationary) Abrikosov lattice, we calculate the intrinsic contribution to $\kappa_{xy}$ using the tight-binding regularization (\ref{eq:Hamiltonian}). We find a new scaling collapse of the extensive numerical results which clearly displays the rapid temperature onset at a small fraction of the maximum pairing gap, followed by an almost linear $T$-dependence.

\section{Model Hamiltonian and density of states scaling}
We work on a two dimensional square lattice of spacing $a$ -- that we set to unity -- and the magnetic field perpendicular to it. Our tight-binding Hamiltonian for the $d$-wave superconductor in the vortex state is
\begin{widetext}
\begin{eqnarray}\label{eq:Hamiltonian}
\mathcal{H}=\sum_{{\bf r}}\left(\sum_{\delta=\hat{x},\hat{y}}\left(t_{{\bf r},{\bf r}+\delta}c^{\dagger}_{{\bf r},\sigma}c_{{\bf r}+\delta,\sigma}+\Delta_{{\bf r},{\bf r}+\delta}\left(c^{\dagger}_{{\bf r},\uparrow}c^{\dagger}_{{\bf r}+\delta,\downarrow}
-c^{\dagger}_{{\bf r},\downarrow}c^{\dagger}_{{\bf r}+\delta,\uparrow}\right)+H.c.\right)-\mu_\sigma c^{\dagger}_{{\bf r},\sigma}c_{{\bf r},\sigma} \right).
\end{eqnarray}
\end{widetext}
This model has been discussed extensively elsewhere\cite{OV2001a,OVAMprl2006,MelikyanTesanovic2006}:
$c_{{\bf r},\sigma}$ is the electron annihilation operator, the sum over the spin projection $\sigma =\uparrow$ or $\downarrow$ in the first and the last terms is implicit, and
$
t_{{\bf r},{\bf r}+\delta}=-te^{-iA_{\bf r,\bf r+\delta}}.
$
The chemical potential and the Zeeman coupling enter via $\mu_{\uparrow(\downarrow)}=\mu\pm h_Z$.
The magnetic flux $\Phi$ through an elementary plaquette enters the Peierls factors as $A_{\bf r,\bf r+\hat{x}}=-\pi y\Phi/\phi_0$, $A_{\bf r,\bf r+\hat{y}}=\pi x\Phi/\phi_0$, where $\phi_0=hc/e$.
The ansatz for the pairing term is
$
\Delta_{{\bf r},{\bf r}+\delta}=\Delta_{\delta} e^{i\theta ({\bf r})}\exp\left(\frac{i}{2}\int_{\bf r}^{\bf r+\delta}d{\bf l}\cdot\nabla\theta\right),
$
with the $d$-wave amplitude $\Delta_{\hat{x}}=-\Delta_{\hat{y}}=\Delta$. Vortex positions ${\bf r}_j$ enter
through $\theta({\bf r})$ which is chosen to be the solution of the (continuum) London's equations
$\nabla\times\nabla \theta({\bf r})=2\pi \hat{z}\sum_{j}\delta({\bf r}-{\bf r}_j)$, $\nabla\cdot\nabla\theta({\bf r})=0$.
The closed form solution for $\theta({{\bf r}})$ can be found in the SI.
Vortices are arranged within a unit cell as shown in the inset of Fig.\ 2: each magnetic unit cell, aligned with the tight-binding lattice, is threaded by magnetic flux $hc/e$ and contains a pair of vortices.  We study a variety of vortex lattices (VL): when $L_x = L_y$ the vortices form a
square lattice, for $L_x / L_y \approx \sqrt 3$ the lattice is triangular, the intermediate ratio $L_x / L_y \approx 1.4$ yields oblique VL.
Performing the singular gauge transformation\cite{FranzTesanovic2000,OV2001a,OVAMprl2006}
turns the hopping and the pairing terms in $\mathcal{H}$ periodic with periodicity $L_x$ and $L_y$, allowing us to use the Bloch theorem\cite{FranzTesanovic2000,OV2001a,OVAMprl2006}.
As such, we perform the particle-hole transformation by letting $\left(c_{\bf r,\uparrow},c^{\dagger}_{\bf r,\downarrow}\right)^T=N^{-1/2}_{uc}\sum_{{\bf k}}e^{i{\bf k}\cdot{\bf r}}
\left(e^{\frac{i}{2}\theta({\bf r})}\psi_{{\bf r,\uparrow}}({\bf k}),e^{-\frac{i}{2}\theta({\bf r})}\psi_{{\bf r,\downarrow}}({\bf k})\right)^T$,
where $\psi_{{\bf r},\sigma}({\bf k})$ is periodic in ${\bf r}\rightarrow {\bf r+R}$ and in the sum $k_{x,y}\in (-\frac{\pi}{L_{x,y}},\frac{\pi}{L_{x,y}}]$.
In terms of the new operators (suppressing ${\bf k}$ on $\psi$'s)
\begin{widetext}
\begin{eqnarray}\label{eq:Hamiltonian2}
\mathcal{H}&=&\sum_{{\bf r}\in u.c.}\sum_{{\bf k}}\left(\sum_{\delta=\hat{x},\hat{y}}\left(
e^{i{\bf k}\cdot\delta}\left(t^{\uparrow\uparrow}_{{\bf r},{\bf r}+\delta}\psi^{\dagger}_{{\bf r},\uparrow}\psi_{{\bf r}+\delta,\uparrow}
-
t^{\downarrow\downarrow}_{{\bf r},{\bf r}+\delta}\psi^{\dagger}_{{\bf r},\downarrow}\psi_{{\bf r}+\delta,\downarrow}\right)+H.c.\right)
-\tilde\mu_{\sigma}\psi^{\dagger}_{{\bf r},\sigma}\psi_{{\bf r},\sigma}\right)
\nonumber\\
&+& \sum_{{\bf r}\in u.c.}\sum_{{\bf k}}\sum_{\delta=\hat{x},\hat{y}}\left(\lambda^{\uparrow\downarrow}_{{\bf r},{\bf r}+\delta}
\left(e^{i{\bf k}\cdot\delta}
\psi^{\dagger}_{{\bf r},\uparrow}\psi_{{\bf r}+\delta,\downarrow}+
e^{-i{\bf k}\cdot\delta}
\psi^{\dagger}_{{\bf r}+\delta,\uparrow}\psi_{{\bf r},\downarrow}
\right)+H.c.\right)
\end{eqnarray}
\end{widetext}
where $t^{\uparrow\uparrow}_{{\bf r},{\bf r}+\delta}={t^{\downarrow\downarrow}}^*_{{\bf r},{\bf r}+\delta} =
t_{{\bf r},{\bf r}+\delta}e^{\frac{i}{2}\theta({{\bf r}+\delta})}e^{-\frac{i}{2}\theta({\bf r})}$, $\lambda^{\uparrow\downarrow}_{{\bf r},{\bf r}+\delta}=\Delta_{{\bf r},{\bf r}+\delta}e^{-\frac{i}{2}\theta({{\bf r}+\delta})}e^{-\frac{i}{2}\theta({{\bf r}})}$, and
$\tilde\mu_{\uparrow(\downarrow)}=h_Z\pm\mu$. Further details, including the treatment of the branch-cuts in $e^{\frac{i}{2}\theta({{\bf r}})}$, are provided in SI.

 Equation (\ref{eq:Hamiltonian2}) has the form of a Hamiltonian for a fictitious {\it semiconductor/metal} with spin-orbit coupling. Although the time-reversal symmetry is explicitly broken in
(\ref{eq:Hamiltonian2}), the vector potential does not couple minimally in all the complex hopping parameters\cite{FranzTesanovic2000}, whether they preserve or flip the spin, and as such there is no Lorentz force on $\psi$'s. By direct analogy with the anomalous Hall effect\cite{NagaosaRMP2010} in semiconductors/metals, any intrinsic $\kappa_{xy}$ is therefore driven by Berry phase mechanisms.

The Heisenberg equation of motion
$i\hbar \frac{\partial}{\partial t} \psi_{\bf r}({\bf k})=\left[\psi_{\bf r}(\bf k),\mathcal{H}\right]=\hat{H}({\bf k})\psi_{\bf r}(\bf k)$ defines the single particle Bloch Hamiltonian, $\hat{H}({\bf k})$, whose discrete eigenvalues, $E_m({\bf k})$, and eigenstates, $|m{\bf k}\rangle$, are labeled by a discrete magnetic sub-band index, $m$.

Clearly, for $\Delta=0$, the model in Eq.(\ref{eq:Hamiltonian}) reduces to the well known square lattice Hofstadter ``butterfly" problem\cite{Hofstadter1976}, with  $\Phi/\phi_0=p/q=L^{-2}_H$ where $L^2_H=L_xL_y$. The Chern integers $\ell_r$, determining the electrical conductivity $\sigma^{(el.)}_{xy}=\frac{e^2}{h}\ell_r$ associated with $M$ filled sub-bands, obey the Diophantine equation\cite{Thouless1982,FradkinBook} $M=s L^2_H+\ell_{r}$, where $s$ is an integer, and where $\ell_r$ is subject to the restriction $|\ell_r|\leq L^2_H/2$. Solving for $\ell_r$ we find that\cite{FradkinBook} (for even $L^2_H$) each of the lowest, and each of the highest, $L^2_H/2-1$ sub-bands carries the Chern number $+1$. In addition, there is a pair of touching bands just above and just below zero energy, which carries the (large) Chern number $2-L^2_H$, equally divided between the two anomalous sub-bands\cite{FradkinBook}.
Near the band extrema the dispersion is parabolic and the energy gap between the magnetic sub-bands is $\hbar\omega_c=4\pi t/L^2_H$.

For ${\bf B}=0$, the quasiparticle spectrum of $\mathcal{H}$ is $E({\bf k})=\pm\sqrt{\epsilon({\bf k})^2+\Delta^2({\bf k})}$,
with $\epsilon({\bf k})=-2t(\cos k_x+\cos k_y)-\mu$ and $\Delta({\bf k})=2\Delta\left(\cos k_x-\cos k_y\right)$.
There are four gapless Dirac points in the $1^{st}$ Brillouin zone, with the velocity anisotropy $\alpha_D=v_F/v_{\Delta}=t/\Delta$, which dominate the low temperature thermodynamics.

Thus, the pairing term in $\mathcal{H}$ mixes $\sim4\Delta/\hbar\omega_c=L^2_H/(\pi \alpha_D)$ sub-bands near the Fermi level. In the physically relevant situation, each of these magnetic sub-bands mixed by $\Delta$ carries a unit Chern number, thus resembling Landau levels of a continuum problem. Moreover, the number of such occupied sub-bands should be large compared to $L^2_H/(\pi \alpha_D)$. Therefore, we choose to set $\mu=2t$; the results presented here are not too sensitive to this choice as long as the pairing term does not mix the anomalous band with the large Chern number, or as long as we are not too close to the band minimum.

\begin{figure}
\includegraphics[width=0.48\textwidth]{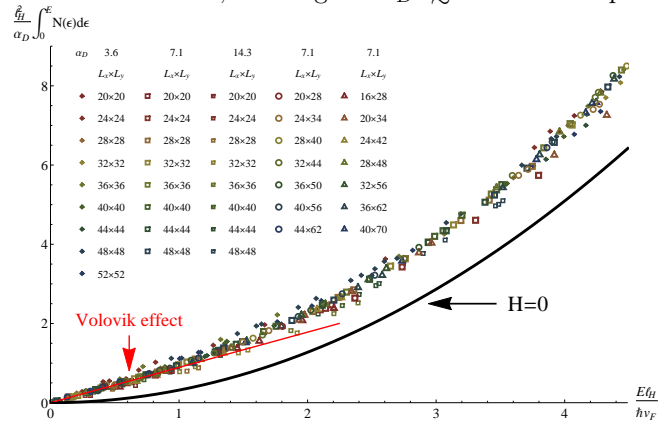}
\caption{
Integrated quasiparticle (qp) density of states (idos) vs. qp energy for a number of magnetic fields, vortex lattice geometries, and Dirac cone anisotropies $\alpha_D=\frac{v_F}{v_\Delta}$ of $3.6$, $7.1$ and $14.3$. The y-axis is rescaled by the square of the magnetic length $\ell^2_H$ and the energy axis $E$ by $\hbar v_F/\ell_H$, where $v_F$ is the Fermi velocity at the Dirac node. The solid curve is the idos for $H=0$ which grows as $\sim E^2$ for small $E$. The Volovik effect corresponds to idos vanishing linearly with $E$ (solid red line).}
\end{figure}

We now turn to ${\bf B}\neq 0$ and $\Delta\neq 0$.  In the semiclassical approximation of Volovik\cite{VolovikJETP1993},
the spatially varying phase $\theta({{\bf r}})$ induces a finite zero-energy density of states, $N(E=0)$, which scales as $\sqrt{H}$. More generally, the scaling expected\cite{SimonLee1997,FranzTesanovic2000} from the approximate Dirac model is $N(E)=\frac{1}{\hbar v_{\Delta}\ell_H}\mathcal{F}'\left(E/\frac{\hbar v_F}{\ell_H},\alpha_D\right)$.
As shown in Fig.2, up to small variations\cite{MelikyanTesanovic2006}, we indeed recover this scaling in our model when $L^2_H/(\pi \alpha_D)\gg 1$.
The integrated density of states per area, per spin, per layer, which is more amenable for comparison when obtained numerically, is seen to follow the scaling
$\int_0^{E}d\epsilon N(\epsilon)=\alpha_D\ell^{-2}_H\mathcal{F}(E/\frac{\hbar v_F}{\ell_H})$, where for $x\ll 1$, $\mathcal{F}(x)=bx$, where ($b\approx0.9$), and for $x\gg 1$, $\mathcal{F}(x)\approx x^2/\pi$, regardless of the shape of the vortex lattice, as long as $\alpha_D\gtrsim 3$.
This implies that the low $T$ specific heat per mol of formula unit is $C/T=0.33\times n_{layers}\times Area[nm^2]\times\frac{\sqrt{H[T]}}{\hbar v_{\Delta}[eV\angstrom]}\frac{mJ}{mol K^2}$. For example, in YBa$_{2}$Cu$_3$O$_{7-\delta}$, $n_{layers}=2$ because there are two CuO$_2$ layers per unit cell whose $Area[nm^2]=0.149$.

\section{Intrinsic contribution to $\kappa_{xy}$}

Having established the connection with known results, we now move to the main focus of this paper.
It has long been known that for ${\bf B}\neq 0$ the thermal transport coefficients cannot be calculated using the Kubo formula alone\cite{Obraztsov1964,Obraztsov1965,SmrckaStreda1977,Cooper1997,OV2001b}. Rather, $\kappa_{xy}$ is given by\cite{OV2001b,SmrckaStreda1977}
\begin{eqnarray}\label{eq:convolution}
\kappa_{xy}=\frac{1}{\hbar T}\int_{-\infty}^{\infty}d\xi \xi^2\left(-\frac{\partial f(\xi)}{\partial\xi}\right)\tilde\sigma_{xy}(\xi) \label{kappaxy_convolution}
\end{eqnarray}
where
$f(\xi)=1/\left(e^{\xi/k_BT}+1\right)$, and
\begin{widetext}
\begin{eqnarray}
\tilde\sigma_{xy}(\xi)
&=&\frac{1}{i}\int\frac{d^2{\bf k}}{(2\pi)^2}\sum_{E_m({\bf k})<\xi<E_n({\bf k})}
\frac{\left\langle m{\bf k} \bigg |\frac{\partial \hat{H}({\bf k})}{\partial k_{x}} \bigg | n{\bf k}\right\rangle
\left\langle n{\bf k} \bigg |\frac{\partial \hat{H}({\bf k})}{\partial k_y} \bigg | m{\bf k}\right\rangle
-(x\leftrightarrow y)
}{\left(E_m({\bf k})-E_n({\bf k})\right)^2}. \label{sigmaxy_definition}
\end{eqnarray}
\end{widetext}
The double sum over $m$ and $n$ is to be performed subject to the stated constrain.
It is well known, that $\tilde\sigma_{xy}(\xi)=\sum_{m}\Omega_m(\xi)$, where $\Omega_m(\xi)$ is the integral over the Berry curvature of the $m^{th}$ magnetic sub-band over the parts of the magnetic Brillouin zone which are below the energy $\xi$. If the sub-band is fully occupied then the integral over the Berry curvature is the first Chern number, i.e., an integer\cite{OV2001b}.

\begin{figure}
\includegraphics[width=0.48\textwidth]{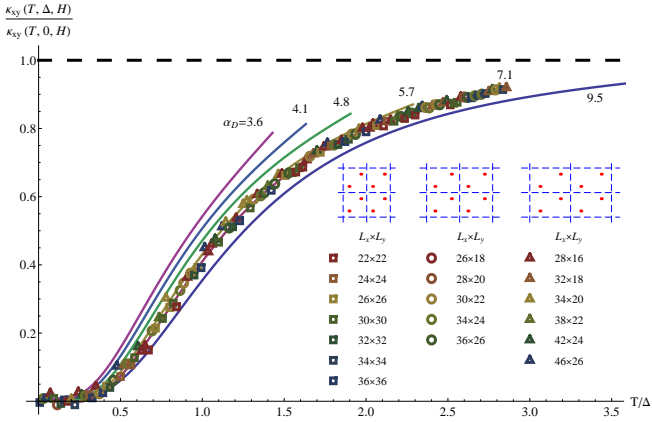} 
\caption{Thermal Hall conductivity $\kappa_{xy}$ in the vortex state of $d_{x^2-y^2}$-wave superconductor vs. the ratio of temperature and the $d$-wave pairing amplitude, $T/\Delta$. The maximum pairing amplitude $\Delta_{max}=4\Delta$. The $y$-axis is rescaled by $\kappa_{xy}(\Delta=0)\sim T/H$. For fixed $\Delta$, the results for varying $H$-fields and vortex lattice geometries (shown in the inset) collapse well onto a single scaling curve, exhibiting a rapid increase at small $T/\Delta$. The scaling functions display an additional weak dependence on the Dirac cone anisotropy, $\alpha_D=v_F/v_{\Delta}=t/\Delta$.}
\end{figure}

Our main result is shown in Fig.\ 2. The ratio $\kappa_{xy}(T,\Delta,H)/\kappa_{xy}(T,0,H)$, in the regime where $\kappa_{xy}(T,0,H)$ is linear in $T$ and $L^2_H/(\pi\alpha_D)\gg 1$, is seen to collapse quite well onto a single curve for each value of the Dirac cone anisotropy $\alpha_D$, and approach $1$ with increasing $T/\Delta$, independent of the vortex lattice geometry.
The rapid onset of $\kappa_{xy}$ with increasing temperature can be understood from the $\xi$-dependence of $\tilde\sigma_{xy}(\xi)$ and the formula in Eq.(\ref{eq:convolution}).
The quantity $\tilde\sigma_{xy}(\xi)$, shown in Fig.\ 3,
has a simple physical interpretation: it is proportional to the quasiparticle contribution to the $T=0$ spin
Hall conductivity\cite{OV2001b} if all the qp sub-bands below the energy $\xi$ are occupied.
In addition, for $\Delta=0$ it can be related to the usual electrical Hall conductivity $\sigma^{(el.)}_{xy}(E_F)$ at the Fermi energy $E_F$,
via $\tilde\sigma_{xy}(\xi)=\frac{h}{e^2}\left(\sigma^{(el.)}_{xy}(\mu+\xi)+\sigma^{(el.)}_{xy}(\mu-\xi)\right)$.
We see that for $\Delta=0$, the quantity $\xi$ can be interpreted as a fictitious Zeeman energy splitting.
Therefore, the loss of the total Chern number in the minority band is largely compensated by the gain in the majority band, and for $\Delta=0$, $\tilde\sigma_{xy}(\xi)$ is essentially independent of $\xi$. For a range of $\xi$'s near zero, its value is approximately $2\ell_r$, i.e., the solution of the mentioned Diophantine equation determining the normal state electrical conductivity at $\mu$ (see $1/\alpha_D$=0 curve in Fig.\ 3 for small $\xi$). This is expected, since an analogous formula to Eq.(\ref{eq:convolution}) holds in the normal state as well\cite{SmrckaStreda1977}, in which case it relates the electrical Hall conductivity to the thermal Hall conductivity.

For $\Delta\neq 0$, the $\tilde\sigma_{xy}(\xi)$ still reaches the value $\sim 2\ell_{r}$, i.e., of order $L^2_H$, when $\xi\gg\Delta$. That is because for $\xi$ large compared to $\Delta$, the contribution to the Chern numbers comes predominantly from the qp bands which are weakly affected by the pairing term.
On the other hand, when $\xi\ll\Delta$, $\tilde\sigma_{xy}(\xi)$ becomes small, non-universal and its value oscillates around zero. This trend
is displayed in Fig.\ 3, where the $\xi$ dependence of $\tilde\sigma_{xy}(\xi)$ is shown for various values of $\Delta$. Such precipitous drop near $\xi=0$ is a consequence of the qp states near zero energy being an almost equal superposition of the electron and a hole, ridding the sub-bands of the Berry curvature. It is fully consistent with the previous result\cite{OVAMprl2006} where the value of $\tilde\sigma_{xy}(0)$, small compared with $L^2_H$, was obtained near $\mu=0$.

\begin{figure}
\includegraphics[width=0.48\textwidth]{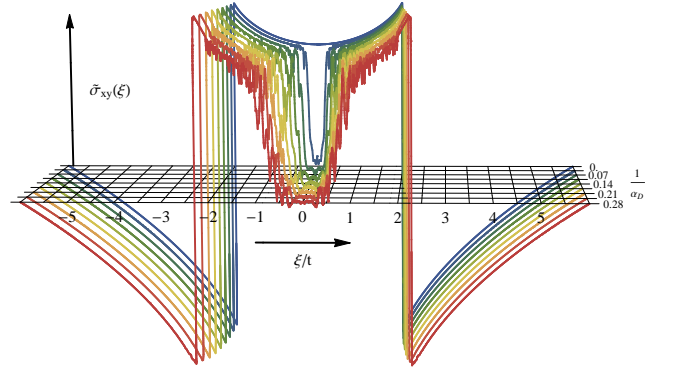} 
\caption{Energy $\xi$ dependence of the $\tilde{\sigma}_{xy}$, which is proportional to the $T=0$ spin Hall conductivity when all the qp bands below $\xi$ are occupied, for various values of the Dirac cone anisotropy $\alpha_D$. The thermal Hall conductivity $\kappa_{xy}$ is obtained from $\tilde{\sigma}_{xy}$ using the equation (\ref{eq:convolution}).}
\end{figure}

Convoluting such $\xi$-dependence of $\tilde\sigma_{xy}(\xi)$ with the thermal factor in Eq.(\ref{eq:convolution}) at low $T$, will result in a vanishingly small $\kappa_{xy}$. As the temperature increases, the thermal function broadens, and the rapid increase of $\tilde\sigma_{xy}(\xi)$ with $\xi$ is mirrored by the rapid increase of $\kappa_{xy}$ with $T$.
In Fig.\ 3, we restrict the $k_BT\ll t$ guaranteeing that in the normal state ($\Delta=0$), $\kappa_{xy}$ is linear in $T$.
The magnetic field dependence follows readily from the above discussion as well. Since for $\xi\gg \Delta$, $\tilde\sigma_{xy}(\xi)\sim L^2_H\sim 1/H$, we find that the $H$-field dependence of $\kappa_{xy}(\Delta\neq 0)$ is entirely determined by the $H$-dependence of $\kappa_{xy}(\Delta=0)\sim 1/H$. This is shown in Fig.\ 3.
We find the dependence of $\kappa_{xy}$ on the Zeeman coupling, $h_Z$, to be barely observable when $h_Z \ll \Delta$.
(see Fig.\ 3 in the SI). This is due to the fact that, with the Zeeman term, we
convolute the average of $\tilde \sigma_{xy}(\xi \pm h_Z)$ instead of $\tilde \sigma_{xy}(\xi)$. For any $h_Z \ll \Delta$,
the main effect of the averaging is to smear the fluctuations in $\tilde \sigma_{xy} (\xi)$, while the overall shape of the curve remains unchanged.
Interestingly, the analogy with the anomalous Hall effect in semiconductors suggests that, in the presence of quenched disorder either in the form of vortex lattice imperfections or impurity scattering, there may be regimes where skew scattering or side jump play a role.

The importance and the novelty of our results stem from the role that the overall $d$-wave pairing strength, $\Delta$, plays in determining the temperature scale at which $\kappa_{xy}$ rapidly increases from a negligibly small value at low $T$. It may therefore provide a means of measuring $\Delta$ in magnetic field via a bulk transport measurement.

\acknowledgments
We wish to thank Prof.\ Ong for sending us their unpublished data. We also acknowledge helpful
conversations with Prof.\ Kun Yang, Dr.\ Ashot Melikyan, and support from the NSF CAREER award (OV) under Grant No.\ DMR-0955561 and
NSF Cooperative Agreement No.\ DMR-0654118 (VC).

\end{document}


\title{Berry phases and the intrinsic thermal Hall effect in high temperature cuprate superconductors \\
  Supplementary Information}
\author{Vladimir Cvetkovic}
\author{Oskar Vafek}
\affiliation{National High Magnetic Field Laboratory and Department
of Physics,\\ Florida State University, Tallahasse, Florida 32306, USA}

\date{\today}

\maketitle

\section {The Hamiltonian and its symmetries}

\begin{figure}
  \centering
    \includegraphics[width=0.75\textwidth]{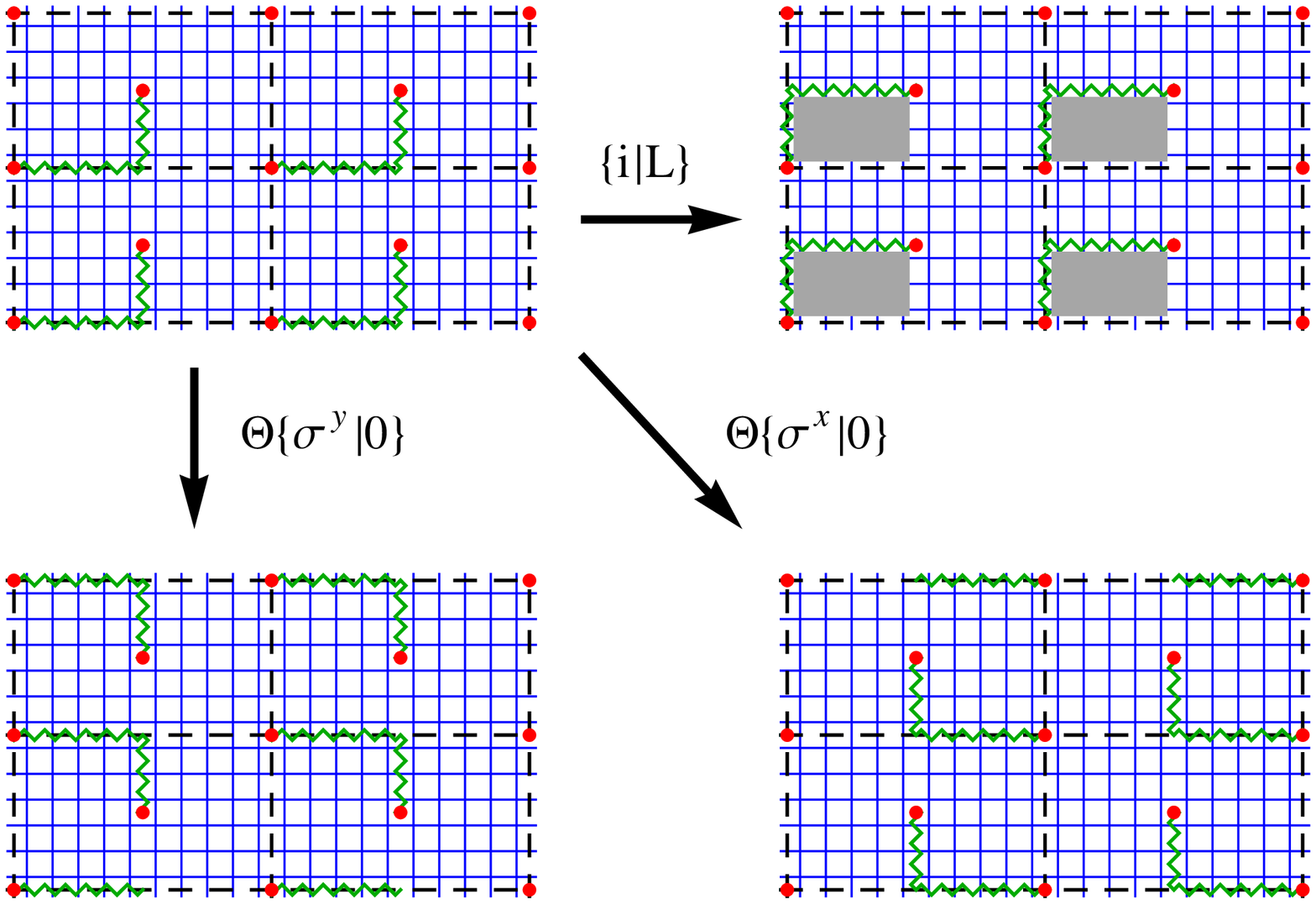}
  \caption{Upper left panel: The original square lattice (light blue), the magnetic unit cell (dashed black lines), the bipartite lattice of vortices (red
    dots), and the branch cuts (green wavy lines). The remaining panels show how the
    branch cut moves under the action of three representative physical symmetries. The gray area in the upper right panel corresponds to sites $\br$
    where $\gamma^\Box_\br = -1$.} \label{FigSupp1}
\end{figure}

We concentrate on the Hamiltonian as presented in Eq.\ (2) in the main text which we restate here:
\begin{eqnarray}
  {\mathcal H} &=& \sum_{\br \in uc} \sum_\bk \left \lbrack \sum_{\delta = \hat x, \hat y} \left ( e^{i \bk \cdot \delta}
    \left ( t_{\br, \br + \delta}^{\uparrow\uparrow} \psi_{\br, \uparrow}^\dagger (\bk) \psi_{\br + \delta, \uparrow} (\bk) + t_{\br, \br + \delta}^{\downarrow\downarrow}
    \psi_{\br, \downarrow}^\dagger (\bk) \psi_{\br + \delta, \downarrow} (\bk) \right ) + h.c. \right ) \right . \nonumber \\
  && \phantom {\sum_{\br \in uc} \sum_\bk} \left .  - \sum_\sigma \tilde \mu_{\sigma} \psi_{\br, \sigma}^\dagger (\bk) \psi_{\br, \sigma} (\bk)
    \right \rbrack \nonumber \\
  &&+ \sum_{\br \in uc} \sum_\bk  \sum_{\delta = \hat x, \hat y} \left ( \lambda_{\br, \br + \delta} \left ( e^{i \bk \cdot \delta} \psi_{\br, \uparrow}^\dagger (\bk)
    \psi_{\br + \delta, \downarrow} (\bk) + e^{-i \bk \cdot \delta} \psi_{\br + \delta}^\dagger (\bk) \psi_{\br, \downarrow} (\bk) \right ) + h.c. \right ). \label{Eq2}
\end{eqnarray}
Here the Franz-Tesanovic singular gauge transformation has been performed \cite{VafekMelikyanFranzTesanovicPRB2001, FranzTesanovicPRL2000,
VafekMelikyanTesanovicPRB2001}, so the the hopping amplitudes are given as
\begin{eqnarray}
  t_{\br, \br + \delta}^{\uparrow \uparrow} = t_{\br, \br + \delta}^{\downarrow \downarrow \ast} &=& - t z_{\br, \br + \delta}^{(2)} e^{i \int_\br^{\br + \delta} \rmd \bl
    \cdot \left ( \frac 12 \nabla \theta - \frac e{\hbar c} \bA \right )}, \label{trr}  \\
  \mu_\sigma &=& \mu + \left \lbrace \begin{array}{l l} +h_Z, & \sigma = \uparrow \\ - h_Z, & \sigma = \downarrow \end{array} \right . , \label{mu} \\
  \lambda_{\br, \br + \delta}^{\uparrow \downarrow} &=& \Delta_\delta z^{(2)}_{\br, \br + \delta}. \label{lambdarr}
\end{eqnarray}
The branch cut variable $z^{(2)}_{\br, \br + \delta}$  is equal to $+1$ on all links except those crossed by the branch cut (green wavy line in the upper left panel of Fig.\ \ref{FigSupp1})
where it is equal to $-1$. The pairing strength $\Delta_{\hat x} = +\Delta$, $\Delta_{\hat y} = -\Delta$ has the $d_{x^2-y^2}$-wave symmetry. Per construction,
$z^{(2)}_{\br, \br + \delta}$ is periodic under the magnetic unit cell translation.

The phase $\theta (\br)$ is a solution to the London equation, $\nabla \times \nabla \theta (\br) = 2 \pi \hat z \sum_j \delta (\br - \br_j)$. With the periodic arrangement
of vortices, as shown in Fig.\ \ref{FigSupp1}, the solution is
\begin{eqnarray}
  \theta (\br) = \sum_j \left ( \arg \sigma (z - z_j; \omega, \omega') + \frac \gamma {2i} \left \lbrack ( z - z_j)^2 - (\bar z - \bar z_j)^2 \right \rbrack + \frac {v_0}{2i} (z \bar z_j - \bar z z_j)
    \right ). \label{theta_sol}
\end{eqnarray}
Here $z = x + i y$, and the summation over $j$ accounts only for the two vortex positions $z_j = x_j + i y_j$. $\sigma (z; \omega, \omega')$ is the
Weierstrass $\sigma$ function with periods $\omega = L_x$ and $\omega' = i L_y$. Constants $\gamma$ and $v_0$ are determined by the boundary
conditions. In order to ensure that the superfluid velocity $\bv \sim \tfrac 1 2 \nabla \theta - \frac e{\hbar c} \bA$ satisfies periodic boundary conditions,
\begin{eqnarray}
  \gamma = \frac 1{L_x} \left ( \frac \pi{2 L_y} - \zeta (\omega; \omega, \omega') \right ) \equiv \frac 1{L_y} \left ( - \frac \pi{2 L_x} + i \zeta (\omega'; \omega, \omega') \right ),
\end{eqnarray}
where $\zeta (\ldots; \omega, \omega')$ is the Weierstrass $\zeta$ function. For square lattices $\gamma = 0$.
Setting $v_0 = \pi / L_H^2$, makes the overall current in the system vanish. The former boundary condition ensures that the hopping amplitudes $t$ in Eq.\ \eqref{Eq2}
are periodic.

The Eq.\ \eqref{Eq2} can be rewritten as
\bea
  {\mathcal H} = \sum_\bk \sum_{\br, \br'} \sum_{\sigma, \sigma'} \psi_{\br, \sigma}^\dagger (\bk) H_{\br, \sigma; \br', \sigma'} (\bk) \psi_{\br', \sigma'} (\bk)
    \equiv \sum_{\bk} \psi^\dagger (\bk) \hat H (\bk) \psi (\bk), \label{H_Bloch}
\eea
defining the single-particle Bloch Hamiltonian $\hat H (\bk)$, which is defined in the main text alternatively through
$i \hbar \frac \partial {\partial t} \psi (\bk) = \lbrack \psi (\bk), {\mathcal H} \rbrack = \hat H (\bk) \psi (\bk)$. Its eigenvalues are $E_m (\bk)$, and the corresponding eigenvectors
$| m; \bk \rangle$.

Due to the specific arrangement of vortices, the Hamiltonian, Eq.\ \eqref{Eq2},
possesses symmetries which consist of either point group operations alone or point group operations combined with the time reversal (TR)
and/or gauge transformations. These symmetries are reflected in the solution for the phase field, Eq.\ \eqref{theta_sol}, and thereby in the
hopping amplitudes Eq.\ \eqref{trr}. From there one can show that the single particle Bloch Hamiltonians at $\hat H (\bk)$ and $\hat H (\bk')$
are related by an unitary or anti-unitary operation, where $\bk'$ and $\bk$ are symmetry related.

In the upper left panel of Fig.\ \ref{FigSupp1} the setup of the Hamiltonian is
presented with the original square lattice drawn in light blue and the boundaries of magnetic unit cells in dashed black. There are two vortices per unit cell
(red dots) connected by branch cuts (green wavy line). The coordinate center coincides with the lower left corner of the magnetic unit cell and the
position of one $A$ vortex, thus ${\bR^A}  = {\bf 0}$. The other vortex in the first magnetic unit cell is at ${\bR^B} = \frac {L_x}2 \hat x + \frac {L_y}2 \hat y$.
For point group operations we use the crystallographic notation, $\{ g | \bt \}$ where $g$ is a point group operation (e.g., mirror) which leaves the coordinate
center intact followed by a shift by $\bt$. In other words, upon a point group operation coordinates transform according to $\{ g | \bt \} \br = g \br + \bt$

One of the transformations which leaves the VL intact is the inversion w.r.t.\ the point halfway between two vortices in a magnetic unit cell,
$\{ i | \bL \}$ where $\bL = \bR^A + \br^B$. Thanks to this symmetry of the VL,
\bea
  e^{i \int_\br^{\br + \delta} \rmd \bl \cdot \left ( \frac 12 \nabla \theta - \frac e{\hbar c} \bA \right )} =
    e^{i \int_{\bL - \br}^{\bL - \br - \delta} \rmd \bl \cdot \left ( \frac 12 \nabla \theta - \frac e{\hbar c} \bA \right )}. \label{eiA_inversion}
\eea
Let us relabel $\psi_{\br, \sigma} (\bk) = \chi_{\bL - \br, \sigma} (-\bk)$ in Eq.\ \eqref{Eq2} so that we can write it as
\begin{eqnarray}
  {\mathcal H} &=& \sum_{\bk' = - \bk} \left \lbrack \sum_{\delta = \hat x, \hat y} \left ( e^{i \bk' \cdot \delta}
    \left ( t_{\bL - \br', \bL - \br' + \delta}^{\uparrow\uparrow} \chi_{\br', \uparrow}^\dagger (\bk') \chi_{\br' + \delta, \uparrow} (\bk') + t_{\bL - \br', \bL - \br' - \delta}^{\downarrow\downarrow}
    \chi_{\br', \downarrow}^\dagger (\bk') \chi_{\br' + \delta, \downarrow} (\bk') \right ) + h.c. \right ) \right . \nonumber \\
  && \phantom {\sum_{\bk' = - \bk}} \left .  - \sum_\sigma \tilde \mu_{\sigma} \chi_{\br', \sigma}^\dagger (\bk') \chi_{\br', \sigma} (\bk')
    \right \rbrack \nonumber \\
  &&+ \sum_{\bk' = - \bk}  \sum_{\delta = \hat x, \hat y} \sum_{\br' = \bL - \br - \delta} \left ( \lambda_{\bL - \br', \bL - \br - \delta} \left ( e^{i \bk' \cdot \delta} \chi_{\br', \uparrow}^\dagger (\bk')
    \chi_{\br' + \delta, \downarrow} (\bk') + e^{-i \bk' \cdot \delta} \chi_{\br' + \delta}^\dagger (\bk') \chi_{\br', \downarrow} (\bk') \right ) + h.c. \right ). \label{Eq2_inversion}
\end{eqnarray}
As implied by the notation, we introduced $\bk' = - \bk$ and $\br' = \bL - \br - \delta$. The term in the first line of Eq.\ \eqref{Eq2_inversion} comes from the ``$h.c.$''
term of the first line in Eq.\ \eqref{Eq2}, and the similar is true for the third line. From Eq.\ \eqref{eiA_inversion} we find that the couplings in Eq.\ \eqref{Eq2_inversion} obey
\begin{eqnarray}
  t_{\bL - \br, \bL - \br - \delta}^{\uparrow \uparrow} = t_{\bL - \br, \bL - \br - \delta}^{\downarrow \downarrow \ast} &=& - t z_{\bL - \br, \bL - \br - \delta}^{(2)}
    e^{i \int_\br^{\br + \delta} \rmd \bl \cdot \left ( \frac 12 \nabla \theta - \frac e{\hbar c} \bA \right )}, \label{trr_inversion}  \\
  \lambda_{\bL - \br, \bL - \br - \delta}^{\uparrow \downarrow} &=& \Delta_\delta z^{(2)}_{\bL - \br, \bL - \br - \delta}. \label{lambdarr_inversion}
\end{eqnarray}
The Hamiltonian Eq.\ \eqref{Eq2_inversion} is therefore describing the same system as Eq.\ \eqref{Eq2} apart from the branch cut which has
been moved by the inversion operation. This Hamiltonian therefore corresponds to the situation in the upper right panel on Fig.\ \ref{FigSupp1}.
The branch cut may be restored to its original position by a gauge transformation, $\gamma^\Box$, which changes sign across the border of
the rectangular area delineated by the original and the new branch cut and shown in gray in Fig.\ \ref{FigSupp1},
\bea
  \gamma^\Box_\br \chi_{\br, \sigma} (\bk) = \chi_{\br, \sigma} (\bk) \left \lbrace \begin {array}{l l} -1, & \br \in \Box \\ +1, & \br \notin \Box \end{array} \right . .
\eea
$\br \in \Box$ means that site $\br$ is inside the rectangular area. After the gauge transformation Eq.\ \eqref{Eq2_inversion} becomes
\bea
  {\mathcal H} &=& \sum_\bk \left ( \gamma^\Box \chi (\bk) \right )^\dagger \hat H (\bk) \left ( \gamma^\Box \chi (\bk) \right ). \label{Eq2_chi_inversion}
\eea

This procedure therefore defines a unitary transformation corresponding to $\{ i | \bL \}$ followed by gauge transformation $\gamma^\Box$,
\bea
  \gamma^\Box \chi (\bk) = U \left ( \gamma^\Box \{ i | \bL \} \right ) \psi (-\bk),
\eea
whose matrix elements are
\bea
  U_{\br, \sigma; \br', \sigma'} \left ( \gamma^\Box \{ i | \bL \} \right ) = \gamma^\Box_\br \delta_{\br, \bL - \br'} \delta_{\sigma, \sigma'}.
\eea
The Kronecker delta $\delta_{\br, \bL - \br'}$ is defined modulo magnetic unit cell translations. Eq.\ \eqref{Eq2_chi_inversion} is then written as
\bea
  {\mathcal H} &=& \sum_\bk \left ( U \left ( \gamma^\Box \{ i | \bL \} \right ) \psi (-\bk) \right )^\dagger \hat H (\bk) \left ( U \left ( \gamma^\Box \{ i | \bL \} \right ) \psi (-\bk) \right ) \nonumber \\
  &=& \sum_{\bk} \psi^\dagger (-\bk) U^\dagger \left ( \gamma^\Box \{ i | \bL \} \right ) \hat H (\bk) U \left ( \gamma^\Box \{ i | \bL \} \right ) \psi (-\bk).
\eea
Comparing the second line with Eq.\ \eqref{H_Bloch} we prove
that $\hat H(-\bk) = U^\dagger \left ( \gamma^\Box \{ i | \bL \} \right ) \hat H (\bk) U \left ( \gamma^\Box \{ i | \bL \} \right )$ which was our aim.

Another operation preserving the vortex arrangement is vertical mirror perpendicular to the $y$-axis and passing through the coordinate center, $\{ \sigma^y | {\bf 0} \}$.
When applied to Eq.\ \eqref{theta_sol} this operation implies
\bea
  e^{i \int_\br^{\br + \delta} \rmd \bl \cdot \left ( \frac 12 \nabla \theta - \frac e{\hbar c} \bA \right )} =
    e^{-i \int_{\sigma^y \br}^{\sigma^y (\br - \delta)} \rmd \bl \cdot \left ( \frac 12 \nabla \theta - \frac e{\hbar c} \bA \right )}. \label{eiA_sigmay}
\eea
The negative sign in the exponent on the right hand side reflects the fact that a vertical mirror reverses the sign of the magnetic field, therefore
we should follow $\{ \sigma^y | {\bf 0} \}$ by the time reversal (TR) operation $\Theta$ which acts as a conjugation. We then relabel
\bea
  \psi_{\br, \sigma} (\bk) = \chi_{\sigma^y \br, \sigma} (-\sigma^y \bk)^\ast = \chi_{\sigma^y \br, \sigma} (\sigma^x \bk)^\ast,
\eea
and repeat the procedure done for the inversion operation. The result is analogous to Eq.\ \eqref{Eq2_inversion}: the Hamiltonian for $\chi (\bk)$ fields 
corresponds to the system with branch cuts repositioned by $\{ \sigma^y | {\bf 0} \}$ operation, as shown in the lower left panel of Fig.\ \ref{FigSupp1}.
A gauge transformation $\gamma^x$ intended to restore the branch cut to its original position must change sign on the vertical line defined by the old and new
portions of the branch cut. For example, it could multiply $\chi_{\br, \sigma} (\bk)$ by $-1$ on the sites left of that line and by $+1$ on the sites right of the line.
However, such a gauge transformation is not periodic unless followed by a phase prefactor:
\begin{eqnarray}
  e^{i \frac {\pi \hat x}{L_x} \cdot \br} \gamma^x_\br \chi_{\br, \sigma} (\bk) = \chi_{\br, \sigma} (\bk)
    e^{i \frac {\pi \hat x}{L_x} \cdot \br} \left \lbrace \begin{array}{c l} +1, & 0 < x < \frac {L_x}2 \\ -1, & \frac{L_x} 2 < x < L_x \end{array} \right . . \label{gammax}
\end{eqnarray}
Here, $x = \br \cdot \hat x$ and $0 \le x < L_x$. The anti-unitary transformation following from this set of operation is therefore defined by its matrix elements
\bea
  \sum_{\br', \sigma'} U_{\br, \sigma; \br', \sigma'} \left ( \gamma^x \{ \sigma^y | {\bf 0} \} \right ) =
    \sum_{\br', \sigma'} \gamma^x_\br \delta_{\br, \sigma^x \br'} \delta_{\sigma, \sigma'},
\eea
and it establishes that
\bea
  \hat H (\sigma^x \bk) &=& e^{-i \frac {\pi \hat x}{L_x} \cdot \br} U^\dagger \left ( \gamma^x \{ \sigma^y | {\bf 0} \} \right )
    \hat H (\bk)^\ast U \left ( \gamma^x \{ \sigma^y | {\bf 0} \} \right ) e^{i \frac {\pi \hat x}{L_x} \cdot \br} \nonumber \\
  &=& U^\dagger \left ( \gamma^x \{ \sigma^y | {\bf 0} \} \right ) \hat H (\frac {\pi \hat x}{L_x} + \bk)^\ast U \left ( \gamma^x \{ \sigma^y | {\bf 0} \} \right ).
\eea

Following the same procedure, we show that vertical mirror $\{ \sigma^x | {\bf 0} \}$ followed by a TR and $e^{i \frac {\pi \hat y}{L_x} \cdot \br}$
maps the original problem to the one with the branch cuts shown in the lower right panel of Fig.\ \ref{FigSupp1}. The branch cuts are restored by
a periodic gauge transformation, $e^{-i \frac {\pi \hat y}{L_y} \cdot \br}$, and this defines
another anti-unitary transformation such that
\bea
  \hat H (\sigma^y \bk) &=& e^{-i \frac {\pi \hat y}{L_y} \cdot \br} U^\dagger \left ( \{ \sigma^x | {\bf 0} \} \right )
    \hat H (\bk)^\ast U \left ( \{ \sigma^x | {\bf 0} \} \right ) e^{i \frac {\pi \hat y}{L_y} \cdot \br} \nonumber \\
  &=& U^\dagger \left ( \{ \sigma^x | {\bf 0} \} \right ) \hat H (\frac {\pi \hat y}{L_y} + \bk)^\ast U \left ( \{ \sigma^x | {\bf 0} \} \right ).
\eea

In the case of a VL which is not a square lattice, all other symmetries, eight in total, are generated from the three we have presented. These are enumerated in
Table \ref{TableSymmetries}. When we introduce the methods for the calculation of the spin Hall conductivity, $\tilde \sigma_{xy} (\xi)$, in the next section, we
use these symmetries to prove that the contributions to $\tilde \sigma_{xy} (\xi)$ from two different symmetry related $\bk$'s are equal. This ultimately speeds
up the calculation by reducing the integration domain from the full magnetic BZ to only a small representative fraction thereof.

The special case of the square lattice, where $L_x = L_y$, possesses additional symmetries. A vertical-diagonal mirror which contains  $A$ and $B$ vortices, followed
by a TR, $\Theta \{ \sigma^Y | {\bf 0} \}$, preserves the VL arrangement.  Since such a transformation changes the position of the branch cut, it must be
followed by $\gamma^\Box$ in order to restore the branch cuts. This is not sufficient to recover the original Hamiltonian, due to the nature of the $d_{x^2-y^2}$-wave gap.
This operation changes the sign of the off-diagonal hopping terms, $\lambda$'s, and must be 
combined with a charge gauge transformation which restores the sign on $\lambda$'s, $\gamma^\lambda = e^{i \pi \sigma^3} = i \sigma_3$.
Only then we can define an anti-unitary operation, such that
\bea
  \hat H (\sigma^X \bk) = U^\dagger \left ( \gamma^\lambda \gamma^\Box \{ \sigma^Y | {\bf 0} \} \right ) \hat H (\bk)^\ast
    U \left ( \gamma^\lambda \gamma^\Box \{ \sigma^Y | {\bf 0} \} \right ).
\eea
The combination of this operation with the previously defined eight operations yields
another eight operation and the total of sixteen point group operations for the case of a square VL.

\begin{table}
\begin{tabular}{| >{$}r<{$} | >{$}c<{$} | >{$}c<{$} | >{$}c<{$} || >{$}r<{$} | >{$}c<{$} | >{$}c<{$} | >{$}c<{$} |}
\hline
\lbrack \gamma \Theta \rbrack \{ g | \bt \} & \lbrack K \rbrack U_{\br, \br'; \sigma, \sigma'} (\ldots) & \bk' & &
  \lbrack \gamma \Theta \rbrack \{ g | \bt \} & \lbrack K \rbrack U_{\br, \br'; \sigma, \sigma'} (\ldots) & \bk' & \\
\hline
\{ e | {\bf 0} \} & \bbone_{\br, \br'} \bbone_{\sigma, \sigma'} & \bk & \Box
  & \gamma^\Box_\br \{ i | \bL \} & \gamma^\Box_\br \delta_{\br, \bL - \br'} \bbone_{\sigma, \sigma'} & -\bk & {\color {gray} \blacksquare} \\
\Theta \{ \sigma^x | {\bf 0} \} & \delta_{\br, \sigma^x \br'} \bbone_{\sigma, \sigma'} K & \frac {\pi \hat y}{L_y} + \sigma^y \bk & {\color {lblue} \blacksquare}
  & \gamma^\Box \Theta \{ \sigma^y | \bL \} & \gamma^\Box_\br \delta_{\br, \bL + \sigma^y \br' } & \frac {\pi \hat y}{L_y} + \sigma^x \bk & {\color {dblue} \blacksquare} \\
\gamma^x \Theta \{ \sigma^y | {\bf 0} \} & \gamma^x_\br \delta_{\br, \sigma^y \br} \bbone_{\sigma, \sigma'}  K & \frac {\pi \hat \hat x}{L_x} + \sigma^x \bk & {\color {lred} \blacksquare}
  & \gamma^x \gamma^\Box_\br \Theta \{ \sigma^x | \bL \} & \gamma^x_\br \gamma^\Box_\br \delta_{\br, \bL + \sigma^x \br' } & \frac {\pi \hat x}{L_x} + \sigma^y \bk
    & {\color {dred} \blacksquare} \\
\gamma^x \{ i | {\bf 0} \} & \delta_{\br, -\br'} \bbone_{\sigma, \sigma'} & \frac {\pi \hat x}{L_x} + \frac {\pi \hat y}{L_y} - \bk & {\color {lgreen} \blacksquare}
  & \gamma^x \gamma^\Box \{ e | \bL \} & \gamma^x_\br \gamma^\Box_\br \delta_{\br, \bL + \br'} \bbone_{\sigma, \sigma'} & \frac {\pi \hat x}{L_x} + \frac {\pi \hat y}{L_y} + \bk
    & {\color {dgreen} \blacksquare} \\
\hline
\end{tabular}
\caption{The symmetries of a triangular VL combined with optional gauge transformations ($\gamma$) and/or TR ($\Theta$) which relate the
  single particle Bloch Hamiltonians at different $\bk$ points. The matrix elements of the corresponding (anti-)unitary transformations are
  listed in the next column, $K$ representing the complex conjugation. The next column gives $\bk'$ related to $\bk$ through the operation,
  and the last column the color corresponding to the area of the magnetic BZ in Fig.\ \ref{FigSuppBZ}.}\label{TableSymmetries}
\end{table}

\begin{figure}
  \centering
    \includegraphics[width=0.25\textwidth]{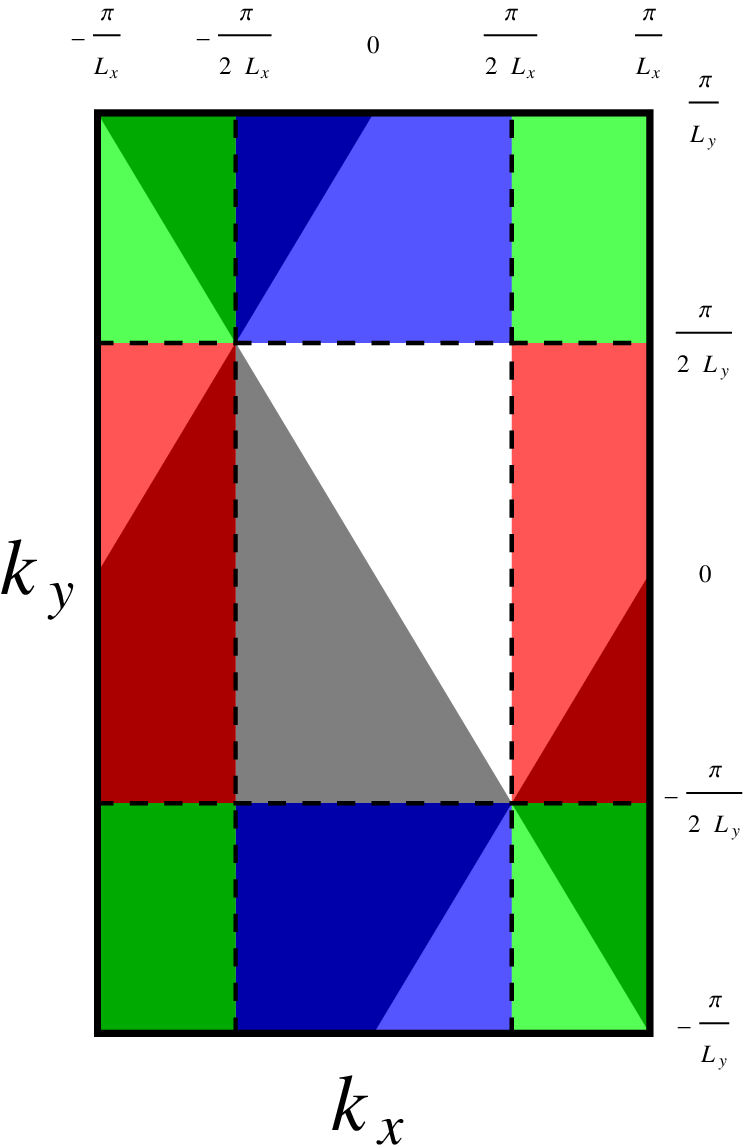}
  \caption{The magnetic Brillouin zone (BZ). The single particle Hamiltonians at $\bk$'s of different colors are related by symmetries tabulated in
    Table \ref{TableSymmetries}. For example, $\bk$ in white and $\bk' = -\bk$ in gray area have their respective single particle Hamiltonians
    related by $\gamma^\Box \{ i | \bL \}$ symmetry.} \label{FigSuppBZ}
\end{figure}

In the absence of the Zeeman term, $h_Z = 0$, another symmetry, which is independent of the VL arrangement, relates
the single particle Hamiltonian at $\bk$ and $-\bk$:
\bea
  \sigma_2 \hat H (\bk) \sigma_2 = - \hat H (-\bk). \label{sigma2symmetry}
\eea
As a result, for each eigenstate $| m; \bk \rangle$ with energy $E_m (\bk)$, there is another state $| m'; -\bk \rangle = \sigma_2 | m; \bk \rangle$ with
energy $E_{m'}(-\bk) = - E_m (\bk)$.

\section{Spin-Hall conductivity}

Thermal Hall conductivity at finite temperature is obtained by convolution of spin Hall conductivity, $\tilde \sigma^s_{\mu \nu} (\xi)$, given
by Eq.\ (3) in the main text which we repeat here:
\begin{eqnarray}
  \kappa_{\mu \nu} (T) = - \frac 1 {\hbar T} \int_{-\infty}^{\infty} \rmd \xi ~  \xi^2 \frac {\rmd f(\xi)}{\rmd \xi} \tilde \sigma^s_{\mu \nu} (\xi). \label{Eq3}
\end{eqnarray}
Here, $f (\xi) = 1 / (e^{\xi / (k_B T)} + 1)$.
The key ingredient in our work is therefore the spin Hall conductivity and here we present methods to calculate it and discuss their efficiency.

In the main text we have the following formula for the transverse spin Hall conductivity (also Ref.\ \onlinecite{VafekMelikyanTesanovicPRB2001}),
\begin{eqnarray}
  \tilde \sigma_{xy}^s (\xi) = \int \frac {\rmd^2 \bk}{(2 \pi)^2} \frac {\hbar^2} i \sum_{E_m (\bk) < \xi < E_n (\bk)}
    \frac {V_x^{mn} (\bk) V_y^{nm} (\bk) - V_y^{mn} (\bk) V_x^{nm} (\bk)}{(E_n (\bk) - E_m (\bk))^2}. \label{sigmaxy_DS}
\end{eqnarray}
The velocity operator is
\begin{eqnarray}
  \hat V_\mu (\bk) = \frac 1 \hbar \frac {\partial \hat H (\bk)}{\partial k_\mu},
\end{eqnarray}
and $V_\mu^{mn} (\bk) = \langle m; \bk | \hat V_\mu (\bk) | n; \bk \rangle$.
Notice that the transverse spin Hall conductivity is
a topological property of the system: $\tilde \sigma_{xy}^s (\xi = 0)$ is proportional to the sum of first Chern number over the occupied bands \cite{VafekMelikyanTesanovicPRB2001}. 

The Eq.\ \eqref{sigmaxy_DS} can be understood as $\tilde \sigma_{xy} (\xi) = \frac 1{2 \pi} \int \rmd^2 \bk ~ {\mathcal F}_\bk (\xi)$, with
\begin{eqnarray}
  {\mathcal F}_\bk (\xi) = \frac {\hbar^2} {2 \pi i} \sum_{E_m (\bk) < \xi < E_n (\bk)} \frac {V_x^{mn} (\bk) V_y^{nm} (\bk) - V_y^{mn} (\bk) V_x^{nm} (\bk)}{(E_n (\bk) - E_m (\bk))^2} \label {F_DS}
\end{eqnarray}
being the Berry curvature summed over all states that are below $\xi$ at $\bk$. We will present several formulas for calculating ${\mathcal F}_\bk (\xi)$ and prove
their equivalence. In the next subsection we will explain the best strategies to calculate $\tilde \sigma_{xy} (\xi)$ based on these formulae.

The double summation over all occupied/unoccupied states in Eq.\ \eqref{F_DS} requires iterating over both $m$ and $n$ which,
when naively implemented, is inefficient and does not have a clear physical connection to the
topological properties of individual bands. This can be addressed by transforming this equation into a sum over individual states with eigenvalues below $\xi$,
\begin{eqnarray}
  {\mathcal F}_\bk (\xi) =  \frac 1{2 \pi i} \sum_{E_m (\bk) < \xi}
    \left ( \left \langle \frac{\partial m_\bk}{\partial k_x} \right | \left . \frac{\partial m_\bk}{\partial k_y} \right \rangle - 
    \left \langle \frac{\partial m_\bk}{\partial k_y} \right | \left . \frac{\partial m_\bk}{\partial k_x} \right \rangle \right ), \label{F_SS}
\end{eqnarray}
where, for brevity, we write $\left | \frac{\partial m_\bk}{\partial k_\mu} \right \rangle = \frac \partial{\partial k_\mu} | m; \bk \rangle$.

When defining the derivatives of the eigenstates in the previous equation, we have to be mindful about the freedom to choose the overall
phase of each individual eigenstate. Transformation $| m; \bk \rangle \to | m; \bk \rangle e^{i \alpha_m (\bk)}$, where
$\alpha_m (\bk)$ is an arbitrary, but smooth, function for each band $m$, leaves Eq.\ \eqref {F_SS} invariant: terms proportional to
$\partial_x \alpha_m (\bk) \partial_y \alpha_m (\bk)$ arise, but mutually cancel. In practice, however, the appearance of such terms leads to
numerical instability of Eq.\ \eqref{F_SS}. In order to evaluate the wave-function derivatives, one needs to diagonalize the Hamiltonian
at $\bk$ as well as in some vicinity of that point. This leads to large terms appearing in this expression. While these are supposed to mutually cancel,
they introduce numerical errors comparable to the value of the integrand.

One way to circumvent the problem is to use projectors in place of the wave functions \cite {AvronSeilerPRL1985}. Projectors,
$P^{(l)} (\bk) = |l; \bk \rangle \langle l; \bk |$, do not carry the ambiguous phase factor, which is
why the previously discussed problem does not occur during the differentiation. Here we derive expressions equivalent to Eq.\ \eqref{F_SS}.
The first step is to evaluate $V_\mu^{mn}$. Differentiating single particle Hamiltonian we obtain
\begin{eqnarray}
  \frac {\partial \hat H (\bk)}{\partial \bk_\mu} = \frac \partial {\partial k_\mu} \sum_l E_l (\bk) P^{(l)} (\bk) = \sum_l \left \lbrack \left ( \frac {\partial E_l (\bk)}{\partial k_\mu} \right ) P^{(l)} (\bk)
    + E_l (\bk) \left ( \frac {\partial P^{(l)} (\bk)}{\partial k_\mu} \right ) \right \rbrack.  \label{partialH}
\end{eqnarray}
From here it follows that
\begin{eqnarray}
  V_\mu^{mn} (\bk) = \frac 1 \hbar \sum_l E_l (\bk) \langle m; \bk | P^{(l)}_{,\mu} (\bk) | n; \bk \rangle, \label{Vmn_partialP}
\end{eqnarray}
assuming that we are interested in terms with $m \neq n$ where the first term from Eq.\ \eqref{partialH} vanishes. We
use the following shorthand notation, $\partial P^{(l)} / \partial k_\mu = P^{(l)}_{,\mu}$. We also drop $\bk$ in the notation from now on.  A useful relation at this step, and later in 
the derivation, follows from differentiating $ P^{(l)} P^{(l')} = \delta_{l l'} P^{(l)}$ identity:
\begin{eqnarray}
  P^{(l)}_{,\mu} P^{(l')} + P^{(l)}  P^{(l')}_{,\mu} = \delta_{l l'} P^{(l)}_{,\mu}.
\end{eqnarray}
Setting $l = l'$, we write Eq.\ \eqref{Vmn_partialP} as
\begin{align}
  V_\mu^{mn} &= \frac 1 \hbar \sum_l E_l \langle m | \left ( P^{(l)} P_{,\mu}^{(l)} + P_{,\mu}^{(l)} P^{(l)} \right ) | n \rangle
    = \frac 1 \hbar \left ( E_m \langle m | P_{,\mu}^{(m)} | n \rangle + E_n \langle m | P_{,\mu}^{(n)} | n \rangle \right ), \label{Vmn_2term}
\end{align}
eliminating the summation over $l$. Differentiating the identity resolution, we obtain another useful identity
\begin{eqnarray}
  \sum_l P^{(l)}_{,\mu} = 0,
\end{eqnarray}
which allows us to write the second term in Eq.\ \eqref{Vmn_2term} as
\begin{align}
  \langle m | P^{(n)}_{,\mu} | n \rangle &= - \langle m | \sum_{l' \neq n} P^{(l')}_{,\mu} | n \rangle
    = - \langle m | \sum_{l' \neq n} \left ( P^{(l')} P_{,\mu}^{(l')} + P_{,\mu}^{(l')} P^{(l')} \right ) | n \rangle
  = - \langle m | P_{,\mu}^{(m)} | n \rangle,
\end{align}
and
\begin{eqnarray}
  V_\mu^{mn} = \frac 1 \hbar ( E_m - E_n ) \langle m | P_{,\mu}^{(m)} | n \rangle.
\end{eqnarray}

The Berry curvature now becomes
\begin{align}
  {\mathcal F}_\bk (\xi) = \frac 1{2 \pi i} \sum_{E_m < \xi < E_n} & \left \lbrack \langle m | P_{,x}^{(m)} | n \rangle
    \langle n | P_{,y}^{(m)} | m \rangle  -  \langle m | P_{,y}^{(m)} | n \rangle \langle n | P_{,x}^{(m)} | m \rangle  \right \rbrack . 
\end{align}
The summation over states with $E_n > \xi$ can be substituted by a summation over states below $\xi$ using the identity resolution,
\begin{eqnarray}
  \sum_{\xi < E_n} | n \rangle \langle n | = \bbone - \sum_{E_m < \xi} | m \rangle \langle m |.
\end{eqnarray}
This yields
\begin{eqnarray}
  {\mathcal F}^{xy} (\xi) = \frac 1{2 \pi i} \sum_\bk \sum_{E_m < \xi} \langle m | \left \lbrack P_{,x}^{(m)}, P_{,y}^{(m)} \right \rbrack | m \rangle
    - \frac 1{2 \pi i} \sum_\bk \sum_{E_m, E_{m'} < \xi} \langle m | \left ( P_{,x}^{(m)} P^{(m')} P_{,y}^{(m)} - P_{,y}^{(m)} P^{(m')} P_{,x}^{(m)} \right ) | m \rangle.
    \label{F_2term}
\end{eqnarray}
Let us first show that the second term is zero. Notice that this term can be cast as 
\begin{align}
  \frac 1{2 \pi i}
  \sum_{E_m, E_{m'} < \xi} \tr \left \lbrack P^{(m)} P^{(m)}_{,x} P^{(m')} P^{(m)}_{,y} - P^{(m)} P^{(m)}_{,y} P^{(m')} P^{(m)}_{,x} \right \rbrack.
\end{align}
Terms in the sum with $m = m'$ vanish due to the cyclical property of the trace.  We are left only with terms where $m \neq m'$. For those,
we use the properties of projector operators and make the following transformations
\begin{align}
  &\frac 1{2 \pi i} \sum_{m \neq m'} \tr \left \lbrack P^{(m)} P^{(m)}_{,x} P^{(m')} P^{(m)}_{,y} - P^{(m)} P^{(m)}_{,y} P^{(m')} P^{(m)}_{,x} \right \rbrack \nonumber \\
  = & \frac 1{2 \pi i} \sum_{m \neq m'} \tr \left \lbrack P^{(m)} P^{(m)}_{,x} P^{(m')} P^{(m')} P^{(m')} P^{(m)}_{,y}
    - P^{(m)} P^{(m)}_{,y} P^{(m')} P^{(m')} P^{(m')} P^{(m)}_{,x} \right \rbrack \nonumber \\
  = & \frac 1{2 \pi i} \sum_{m \neq m'} \tr \left \lbrack P^{(m)} P^{(m)} P^{(m')}_{,x} P^{(m')} P^{(m')}_{,y} P^{(m)}
    - P^{(m)} P^{(m)} P^{(m')}_{,y} P^{(m')} P^{(m')}_{,x} P^{(m)} \right \rbrack \nonumber \\
  = & \frac 1{2 \pi i} \sum_{m \neq m'} \tr \left \lbrack P^{(m')} P^{(m')}_{,y} P^{(m)} P^{(m')}_{,x} - P^{(m')} P^{(m')}_{,x} P^{(m)} P^{(m')}_{,y} \right \rbrack .
\end{align}
Notice that the first and the last line are equal up to a minus sign, therefore this expression must be equal to zero.

We are left with the first term in Eq.\ \eqref{F_2term}, that is,
\begin{eqnarray}
  {\mathcal F}_\bk (\xi) = \frac 1{2 \pi i} \sum_{E_m < \xi} \tr \left ( P^{(m)} \left \lbrack P_{,x}^{(m)}, P_{,y}^{(m)} \right \rbrack \right ). \label{F_PdPdPm}
\end{eqnarray}
Each term in the sum is the Berry curvature of $m$'th band at point $\bk$ in the BZ.
An equivalent formula can be found in Ref.\ \onlinecite{AvronSeilerPRL1985}.

While Eq.\ \eqref{F_PdPdPm} does not suffer from the numerical instability the same way Eq.\ \eqref{F_SS}, we found that it may become unstable
at, and in the vicinity of, $\bk$'s where two bands are degenerate. When this is the case there is an SU(2) freedom in choosing the respective projectors
for the two degenerate bands at $\bk$. Just like before, this freedom results in arbitrarily large terms that are supposed to cancel, but instead
introduce an error of the order of ${\mathcal F}_\bk (\xi)$. We found an alternative, but related, numerically more stable expression to
remedy this problem. Instead of calculating Berry curvature of individual bands, here we define projector onto all states below energy $\xi$,
\begin{eqnarray}
  P (\xi) = \sum_{E_m < \xi} P^{(m)}.
\end{eqnarray}
Utilizing the identities for projector operators show above, it can be shown that Eq.\ \eqref{F_PdPdPm} is equivalent to
\begin{eqnarray}
  {\mathcal F}_\bk (\xi) = \frac 1{2 \pi i} \tr \left ( P ( \xi) \left \lbrack P_{,x}(\xi), P_{,y}(\xi) \right \rbrack \right ).\label{F_PdPdP}
\end{eqnarray} 
This expression does not suffer from any of the previously discussed numerical instabilities.

\subsection{Strategies for computing ${\mathcal F}_\bk (\xi)$}

Calculating the spin Hall conductivity, translates into doing a Riemann sum over the Brillouin zone (BZ) with ${\mathcal F}_\bk (\xi)$ as
the integrand. The method we
developed for the Riemann summation is presented in the next section. Here
we concentrate on the methods for calculating ${\mathcal F}_\bk (\xi)$ and analyze their computational complexity.
The calculation of ${\mathcal F}_\bk (\xi)$ requires computationally costly operations, diagonalization of large matrices (dimension $2 L_H^2$),
construction of projectors, summation over
a large number of states, etc. Here we outline our algorithm for performing  these operations in the most efficient way,
using either Eq.\ \eqref{sigmaxy_DS}, Eq.\ \eqref{F_PdPdPm}, or Eq.\ \eqref{F_PdPdP}

The convolution in Eq.\ \eqref{Eq3} requires the knowledge of $\tilde \sigma_{xy} (\xi)$ at arbitrary values of $\xi$; therefore it is desirable that we are able
to return the value of ${\mathcal F}_\bk (\xi)$ quickly for any $\xi$. An important observation is that ${\mathcal F}_\bk (\xi)$ is a constant as a function of
$\xi$ unless $\xi$ coincides with one of the Hamiltonian eigenvalues $E_l$; there ${\mathcal F}_\bk (\xi)$ exhibits a jump. This is true regardless of which
definition of ${\mathcal F}_\bk (\xi)$ we use. 
We rely on this property to create an ordered lookup table where we store pairs of values $(E_l, {\mathcal F}_\bk (E_l + 0^+))$, i.e., the list of eigenvalues and ${\mathcal F}_\bk$ at
$\xi$'s just above each eigenvalue. Once the lookup table has been created, finding ${\mathcal F}_\bk (\xi)$ requires a search for the
highest $E_l < \xi$ in the lookup table and reading off the corresponding ${\mathcal F}$ (when $\xi$ is below the bottom
of the band ${\mathcal F}_\bk (\xi) = 0$). Given that the table is ordered, the complexity of such an operation is $O(\log L_H)$.

What about the computational complexity of the lookup table generation? The diagonalization of the Hamiltonian at $\bk$ is the first step regardless
the method of finding ${\mathcal F}_\bk (\xi)$. Its computational cost is $O (L_H^6)$ as we need the entire spectrum as well as all the eigenstates.
In the case when we are supposed to find the derivatives of the projectors, we need to perform several diagonalizations in the vicinity of the $\bk$ point
and find the projectors there to. This adds an overall constant factor to the computational cost, but does not change its asymptotic behaviour.

For each eigenvalue $E_l$, building the corresponding projector $P (E_l + 0^+)$ requires computation of each element of the projector matrix, which is a
dense matrix, thus $O (L_H^4)$
operations. While it may seem that for $E_l$'s far from both the bottom and the top of the band one needs to repeat this $l \sim O(L_H^2)$ times, once for each band
below $E_l$, this can be avoided by deriving $P (E_l + 0^+)$ from already constructed $P (E_{l - 1} + 0^+)$,
\begin{eqnarray}
  P (E_l + 0^+) = P (E_{l-1} + 0^+) + | l \rangle \langle l |.
\end{eqnarray}
Calculating the derivatives, as mentioned before, requires this operation to be repeated several times in the vicinity of $\bk$. While adding a constant prefactor,
these operations do not change the computational complexity of this step asymptotically. The most computationally expensive step comes from evaluating
the product of matrices in Eq.\ \eqref{F_PdPdP}. Naive matrix product operation has a computational complexity proportional to the cube of the
matrix size, in our case $O(L_H^6)$. Having to repeat this for each entry in the lookup table, we conclude that the construction of the lookup
table for ${\mathcal F}_\bk (\xi)$ has the computational complexity which scales as $O (L_H^8)$. Methods for matrix multiplication that
scale with lower power of the matrix size \cite{Strassen1969, CoppersmithWinograd1990} are known. Overlooking the fact that these algorithms offer only a
negligible ($\lesssim 15\%$) speedup over the standard library methods \cite{DAlbertoNicolau2008} for typical matrix sizes we use,
we note that even the most sophisticated matrix-product algorithm presently known \cite{LeGall2014} cannot reduce the asymptotic computational complexity of
the lookup table creating below $O(L_H^{6.745})$. The theoretical limit on the matrix multiplication complexity \cite{Raz2002} sets the
lower bound for the lookup table creation with projectors to $O(L_H^6 \log L_H)$. We conclude that this is the lower bound for the computational 
complexity of the lookup tables for ${\mathcal F}_\bk (\xi)$ based on either Eq.\ \eqref{F_PdPdPm} and Eq.\ \eqref{F_PdPdP}. In practice, however,
we found that the execution time of the lookup table computation scaled as $O(L_H^{8 - \epsilon})$ with $\epsilon < 1$.

Here we show that we can do much better using the original, double sum, formula, Eq.\ \eqref{F_DS}. Naively,
the integrand ${\mathcal F}_\bk (\xi)$ in that equation contains summation over two indices, $m$ and $n$, and both sums have $O(L_H^2)$ terms
unless $\xi$ is close to the bottom or the top of the band. Further, finding each $V_\mu^{mn}$ for a general $V_\mu$
matrix would require summing $O(L_H^4)$ terms. In total, these estimates imply that the creation of a lookup table
has  $O(L_H^{10})$ computational complexity. We reduce the complexity to $O(L_H^6)$ using the following observations.

First notice that the difference between two adjacent ${\mathcal F}$'s in the lookup table is given by
\begin{eqnarray}
  {\mathcal F}_\bk (E_l + 0^+) - {\mathcal F}_\bk (E_{l-1} + 0^+) &=& \frac {\hbar^2}{2 \pi i}
    \sum_{m=1}^{l} \sum_{n = l+1}^{2 L_H^2} \frac {V_x^{mn} V_y^{nm} - V_y^{mn} V_x^{nm}}{(E_n - E_m)^2} -  \frac  {\hbar^2}{2 \pi i}
    \sum_{m=1}^{l-1} \sum_{n = l}^{2 L_H^2} \frac {V_x^{mn} V_y^{nm} - V_y^{mn} V_x^{nm}}{(E_n - E_m)^2} \nonumber \\
  &=& \frac  {\hbar^2} {2 \pi i} \sum_{n = l+1}^{2 L_H^2} \frac {V_x^{ln} V_y^{nl} - V_y^{ln} V_x^{nl}}{(E_n - E_l)^2}
     -  \frac  {\hbar^2}{2 \pi i} \sum_{m=1}^{l-1} \frac {V_x^{ml} V_y^{lm} - V_y^{ml} V_x^{lm}}{(E_l - E_m)^2} \nonumber \\
  &=&  \frac  {\hbar^2}{2 \pi i}  \sum_{\substack{m=1 \\ m \neq l}}^{2 L_H^2} \frac {V_x^{ml} V_y^{lm} -  V_y^{ml} V_x^{lm}}{(E_l - E_m)^2} . \label{Fdiff_DS}
\end{eqnarray}
The most of the terms in the difference cancel and we are left with only $2 L_H^2 - 1$ terms to sum over. 
Therefore, if we build the lookup table by adding Eq.\ \eqref {Fdiff_DS} to ${\mathcal F}_\bk (E_{l-1} + 0^+)$ to
obtain  ${\mathcal F}_\bk (E_{l} + 0^+)$,  for each lookup table entry we sum only $O(L_H^2)$ terms instead of $O(L_H^4)$. This
leads to $O(L_H^2)$ improvement over the original double summation formula, Eq.\ \eqref{F_DS}.

Next, we observe that $V_\mu$ are sparse matrices, each having only $O(L_H^2)$ non-zero elements. This follows from the fact
that they derive from the single particle Hamiltonian $\hat H (\bk)$ which is
itself a sparse matrix. The computation of $V_\mu^{mn}$ therefore requires a summation of only $O(L_H^2)$ terms, thereby improving the computational
time by another $O(L_H^2)$ factor. The computation of $V_\mu^{mn}$'s can further be optimized by noticing that, once $l$ is fixed,
we can find $V_\mu | l \rangle$, store those two vectors, and use them to compute $V_\mu^{ml}$ and $V_\mu^{ln}$ faster.

Overall, we find that the computational complexity of the lookup table creation is $O(L_H^6)$ if Eq.\ \eqref{sigmaxy_DS} is used in
conjunction with these improvements. A further advantage of this method over those using the projectors is that, for each ${\mathcal F}_\bk$,
the diagonalization of the Hamiltonian has to be performed only once, at $\bk$ point.

\subsection {Symmetries and ${\mathcal F}_\bk (\xi)$}

Now we show that, if two Bloch Hamiltonians, $\hat H (\bk)$ and $\hat H (\bk')$ are related by an unitary or anti-unitary transformation, then
${\mathcal F}_{\bk} (\xi) = {\mathcal F}_{\bk'} (\xi)$ for all $\xi$. This property allows us to calculate the contributions to $\tilde \sigma_{xy} (\xi)$
coming only from a fraction of the magnetic BZ which contains symmetry inequivalent $\bk$ points only; the contributions from the
remainder of the integration domain is then proportional to what we have calculated. In Fig.\ \ref{FigSuppBZ} such an area is represented by a
white triangle for the case of a non-square VL. Since this area is one eight of the original magnetic BZ, the use of symmetries speeds up our calculation by a 
factor of 8. When the vortices form a square lattice, the integration domain is further reduced by a factor of two and the speed of the calculation is
correspondingly improved.

Let us consider, for illustration, $\gamma^\Box \{ i | \bL \}$ operation in Table \ref{TableSymmetries} for which $\bk' = -\bk$. We have
$| m; -\bk \rangle = U^\dagger \left ( \gamma^\Box \{ i | \bL \} \right ) | m; \bk \rangle = U^\dagger | m; \bk \rangle$
with $U$ listed in the table and its argument suppressed. We also have, $\hat H ({-\bk}) = U^\dagger \hat H (\bk) U$. Since $U$ is $\bk$-independent we have
\begin{eqnarray}
  \hat V_\mu (-\bk) = \left . \frac \partial{\partial \bq_\mu} \hat H ({\bq}) \right |_{\bq \to -\bk} = - U^\dagger \hat V_{\mu} (\bk) U.
\end{eqnarray}
Here we used the fact that the derivative w.r.t.\ $\bq$  of an even function in $\bq$ is an odd function. Combining these identities we find
that Eq.\ \eqref{F_DS}, evaluated at $-\bk$, becomes
\begin{eqnarray}
  {\mathcal F}_{-\bk} (\xi) &=& \frac {\hbar^2} {2 \pi i} \sum_{E_m < \xi < E_n}
    \frac { \langle m; -\bk | \hat V_x (- \bk) | n; -\bk \rangle \langle n; -\bk | \hat V_y (-\bk) | m; -\bk \rangle - h.c.}{(E_n - E_m)^2} \nonumber \\
  &=& \frac {\hbar^2} {2 \pi i} \sum_{E_m < \xi < E_n} \frac { \langle m; \bk | U U^\dagger \hat V_x (- \bk) U U^\dagger | n; \bk \rangle \langle n; \bk |
     U U^\dagger \hat V_y (-\bk) U U^\dagger | m; \bk \rangle - h.c.}{(E_n - E_m)^2} \nonumber \\
  &=& \frac {\hbar^2} {2 \pi i} \sum_{E_m < \xi < E_n} \frac { \langle m; \bk | \hat V_x (\bk) | n; \bk \rangle \langle n; \bk | \hat V_y (\bk) | m; \bk \rangle - h.c.}{(E_n - E_m)^2}
  = {\mathcal F}_{\bk} (\xi)
\end{eqnarray}
We used $E_l (-\bk) = E_l (\bk)$ and dropped the $\bk$-argument accordingly.  The proof is similar for other unitary symmetry operations.

To illustrate how to equate ${\mathcal F}_{\bk} (\xi)$ for $\bk$ and $\bk'$ related by an anti-unitary  symmetry, let us consider $\Theta \{ \sigma^x | {\bf 0} \}$
in Table \ref{TableSymmetries}. For this operation $\bk' = \frac {\pi \hat y}{L_y} + \sigma^y \bk$ and we use $\bk'$ for brevity. Since this operation is anti-unitary,
$| m; \bk' \rangle = U^\ast | m; \bk \rangle^\ast$. As before, $\hat H ({\bk'}) = U^\dagger \hat H ({\bk})^\ast U$ which implies
$\hat V_x (\bk') = - U^\dagger \hat V_x (\bk)^\ast U$ and $\hat V_y (\bk') = U^\dagger \hat V_y (\bk)^\ast U$. Combining these we find
\begin{eqnarray}
  {\mathcal F}_{\bk'} (\xi) &=& \frac {\hbar^2} {2 \pi i} \sum_{E_m < \xi < E_n}
    \frac { \langle m; \bk' | \hat V_x (\bk') | n; \bk' \rangle \langle n; \bk' | \hat V_y (\bk') | m; \bk' \rangle - h.c.}{(E_n - E_m)^2} \nonumber \\
  &=& \frac {\hbar^2} {2 \pi i} \sum_{E_m < \xi < E_n} \frac { \langle m; \bk | U U^\dagger \hat V_x (\bk) U U^\dagger | n; \bk \rangle^\ast \langle n; \bk |
     U U^\dagger (- \hat V_y (\bk)) U U^\dagger | m; \bk \rangle^\ast - h.c.}{(E_n - E_m)^2} \nonumber \\
  &=& \frac {\hbar^2} {2 \pi i} \sum_{E_m < \xi < E_n} \frac { \langle m; \bk | \hat V_x (\bk) | n; \bk \rangle \langle n; \bk | \hat V_y (\bk) | m; \bk \rangle - h.c.}{(E_n - E_m)^2}
  = {\mathcal F}_{\bk} (\xi)
\end{eqnarray}
We have just shown that ${\mathcal F}_{\frac {\pi \hat y}{L_y} + \sigma^y \bk} (\xi) = {\mathcal F}_\bk (\xi)$. The proof is analogous for all other anti-unitary operations.

From here it follows that ${\mathcal F}_{\bk'} (\xi) = {\mathcal F}_\bk (\xi)$  for eight $\bk'$ corresponding to each operation in Table \ref{TableSymmetries}.
In Fig.\ \ref{FigSuppBZ} we colour different sections of the magnetic BZ according to these symmetries. It is sufficient to calculate the Berry curvature in one triangle, 
say the white one, to be able to find the Berry curvature in the entire magnetic BZ by applying these symmetries. Hence, in order to find the spin Hall conductivity we
only need to concentrate on the white triangle and multiply the result we obtain by 8.

In the case of the square VL, the diagonal mirrors and the symmetries generated through these fold the area of the integration further
by a factor of two, since, in this case we can prove that ${\mathcal F}_{(k_x, k_y)} (\xi) = {\mathcal F}_{(k_y, kx)} (\xi)$. When this is the case, the integration domain is one half of the
white triangle, and this domain is defined by corners at $(0, 0)$, $(\pi / 2 \ell_x, \pi / 2 \ell_y)$, and $(\pi / 2 \ell_x, -\pi / 2 \ell_y)$.

The symmetry of the Hamiltonian expressed in Eq.\ \eqref{sigma2symmetry}, since unitary, implies that
\bea
  {\mathcal F}_\bk (E_l + 0^+) - {\mathcal F}_\bk (E_{l-1} + 0^+) = {\mathcal F}_{-\bk} (-E_l - 0^+) - {\mathcal F}_{-\bk} (- E_{l-1} - 0^+), \label{Fk_Fmink}
\eea
where $E_l$'s are the eigenvalues at $\bk$ and, thanks to this symmetry, $-E_l$'s are the eigenvalues at $-\bk$.
Given that ${\mathcal F}_\bk (\xi)$ is a constant unless $\xi$ coincides with an eigenvalue at $\bk$, the right hand side of Eq.\ \eqref{Fk_Fmink},
we can rewrite that as
\bea
  {\mathcal F}_\bk (E_l + 0^+) - {\mathcal F}_\bk (E_{l-1} + 0^+) = - \left ( {\mathcal F}_{-\bk} (-E_{l} + 0^+) - {\mathcal F}_{-\bk} (- E_{l+1} + 0^+) \right ). \label{Fk_Fmink2}
\eea
The expression on the left hand side contributes to $\tilde \sigma_{xy} (\xi = E_l)$, whereas the one on the right hand side contributes to $\tilde \sigma_{xy} (\xi = - E_l)$
up to the minus sign. Therefore, we conclude that $\tilde \sigma_{xy} (\xi)$ is an even function of $\xi$,
\bea
   \tilde \sigma_{xy} (\xi) = \tilde \sigma_{xy} (-\xi).
\eea
When the VL has the inversion 
symmetry, as it is the case for us, this symmetry also implies that ${\mathcal F}_{\bk} (\xi)$ is an even function of $\xi$. We therefore calculate ${\mathcal F}_{\bk} (\xi)$
only for negative $\xi$'s and from there infer the values for positive $\xi$, thereby reducing the computation time by another factor of two.

\subsection {Strategies for computing $\tilde \sigma_{xy} (\xi)$}

Recalling the equation for the spin Hall conductivity from the previous section, $\tilde \sigma_{xy} (\xi) = \frac 1{2 \pi} \int \rmd^2 \bk ~ {\mathcal F}_\bk (\xi)$,
we see that $\tilde \sigma_{xy} (\xi)$ can be approximated and calculated by expressing it as a Riemann sum,
\begin{eqnarray}
  \tilde \sigma_{xy} (\xi) \approx \frac 1{2 \pi} \sum_{i = 1}^s A_i {\mathcal F}_{\bk_i} (\xi), \label{sigmaxyRiemann}
\end{eqnarray}
where the integration domain (in our case a fraction of the magnetic BZ defined previously) is partitioned into $s$ areas, each with area $A_i$. For each
area, the Berry curvature is calculated at some point $\bk_i$ within that area. The partitioning of the integration domain is performed using adaptive mesh
integration method. The details of the method will be presented elsewhere, it suffices to know that this method starts with a coarse mesh
of $\bk$-points and then refines the mesh, i.e., increases the density of $\bk$-points in the areas where the absolute errors are the largest. In this
manner the overall error is minimized using the fewest number, $s$, of $\bk$-points where ${\mathcal F}_\bk (\xi)$ is calculated.

The calculation of ${\mathcal F}_\bk (\xi)$ at $s$ points requires $O(s L_H^6)$ operations using the method presented previously and this is the most
time consuming part of our calculation. From here, we could store the lookup tables for ${\mathcal F}$ at each $\bk_i$ and invoke Eq.\ \eqref{sigmaxyRiemann} whenever 
we need $\tilde \sigma_{xy} (\xi)$. Each such call would require $O (s \log L_H)$ time. For us $10^3 \lesssim s \lesssim 3 \cdot 10^3$, and this method
might slow down the calculation of the spin-Hall conductivity significantly. Therefore, we constructed another lookup table holding values of $\xi$ and the corresponding
$\tilde \sigma_{xy} (\xi)$.

We rely on the fact that all the terms in Eq.\ \eqref{sigmaxyRiemann} are constant as a function of $\xi$, except for $\xi$'w which coincide with any of the
eigenvalues $E_m ({\bk_i})$; there the sum exhibits a discontinuous jump originating only from ${\mathcal F}_{\bk_i} (\xi)$. The lookup table therefore
is an ordered list of pairs $\left ( \xi, \tilde \sigma_{xy} (\xi + 0^+) \right )$. For each eigenvalue $E_m ({\bk_i})$ there is a pair with $\xi = E_m ({\bk_i})$ and
$\tilde \sigma_{xy} (\xi + 0^+)$ is the value of the spin-Hall conductivity at $\xi$ infinitesimally above $\xi$. Such a lookup table contains $2 s L_H^2$. Since it
is ordered, finding $\tilde \sigma_{xy} (\xi)$ for any given $\xi$ takes $O (\log s) + 2 O (\log L_H)$ time which is much faster than the naive implementation
mentioned before.

An naive way to build the lookup table is to get all the eigenvalues  $E_m ({\bk_i})$, and find $\tilde \sigma_{xy} (E_m (\bk_i) + 0^+)$ according to
Eq.\ \eqref{sigmaxyRiemann}. Since there are $2 s L_H^2$ distinct eigenvalues, such an approach has computational complexity $O (s^2 L_H^2 \log L_H)$.
While this calculation step would not be the most time consuming, we nevertheless found a much more efficient algorithm to generate the lookup table
for $\tilde \sigma_{xy} (\xi)$. We first notice that, for any two consecutive eigenvalues, $E_{m} (\bk_j) \le E_{m'} (\bk_{j'})$ (here $\bk_j$ and $\bk_{j'}$ may
or may not be the same), the difference of the spin-Hall conductivities corresponding to these is given by
\begin{eqnarray}
  \tilde \sigma_{xy} (E_{m'} (\bk_{j'}) + 0^+) - \tilde \sigma_{xy} (E_{m} (\bk_j) + 0^+) &=& \frac 1{2 \pi} \sum_{i = 1}^s A_i {\mathcal F}_{\bk_i} (E_{m'} (\bk_{j'}) + 0^+) - 
    \frac 1{2 \pi} \sum_{i = 1}^s A_i {\mathcal F}_{\bk_i} (E_m (\bk_j) + 0^+)  \nonumber \\
  &=& \frac 1{2 \pi} \sum_{i = 1}^s A_i \left ( {\mathcal F}_{\bk_i} (E_{m'} (\bk_{j'}) + 0^+) - {\mathcal F}_{\bk_i} (E_m (\bk_j) + 0^+)  \right ) \nonumber \\
  &=& \frac 1{2\pi} \left ( {\mathcal F}_{\bk_{j'}} (E_{m'} (\bk_{j'}) + 0^+) - {\mathcal F}_{\bk_{j'}} (E_{m' - 1} (\bk_{j'}) + 0^+) \right ). \label{sigmaxy_update}
\end{eqnarray}
Since there are no other eigenvalues between $E_{m} (\bk_j)$ and $E_{m'} (\bk_{j'})$, all ${\mathcal F}_{xy}$'s except ${\mathcal F}_{\bk_{j'}} (\xi)$  are constant 
and therefore cancel. The change between two consecutive $\tilde \sigma_{xy} (\xi)$'s in the lookup table is therefore calculated by updating the contribution
from $\bk_{j'}$ only. This requires only two operations.

We further optimize the processing of all the eigenvalues in the correct, sorted, order by utilizing priority queue, a data structure in which both inserting (pushing) new elements
and returning the smallest one (popping) take time proportional to the logarithm of the number of elements in the structure. We do that by initially building the queue by
pushing the smallest eigenvalue at each $\bk_i$. In each iteration the smallest eigenvalue, $E_m ({\bk_i})$, in the queue is popped and replaced by the next eigenvalue
at the same $\bk_i$ (until the eigenvalues are exhausted). Since the eigenvalues at each $\bk_i$ are sorted, we are guaranteed to pop all the
eigenvalues in the ascending order. Whenever an eigenvalue is popped, we find the spin-Hall conductivity $\tilde \sigma_{xy} (E_m ({\bk_i}) + 0^+)$ according
to Eq.\ \eqref{sigmaxy_update} and add a new entry to the lookup table. Since each iteration has computational complexity $O (\log s)$, the total computational complexity
for building the entire lookup table for $\tilde \sigma_{xy} (\xi)$ using this algorithm is $O (s L_H^2 \log s)$.

\section {$\kappa_{xy}$ in the presence of Zeeman effect}

\begin{figure}
  \centering
    \includegraphics[width=0.6\textwidth]{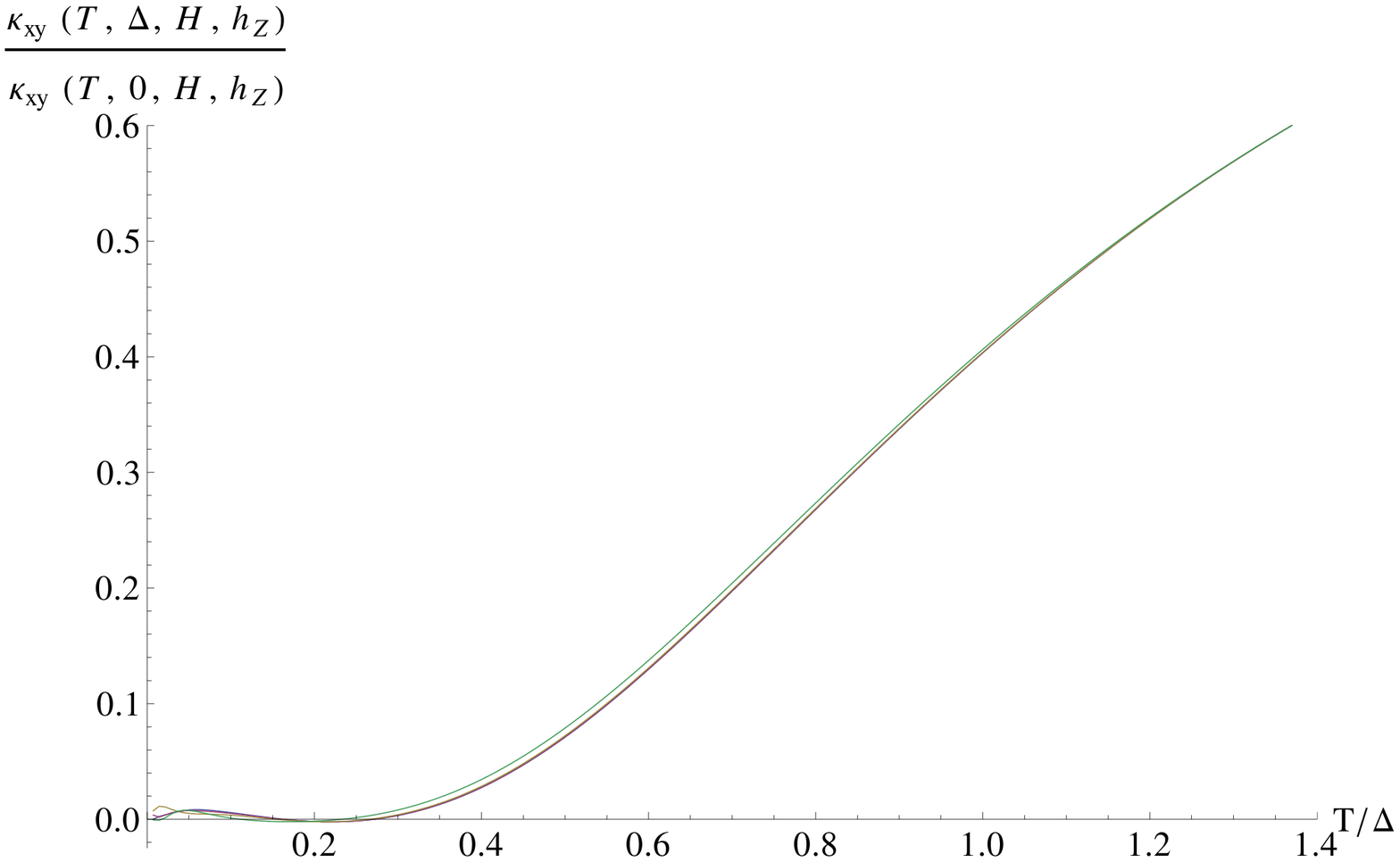}
  \caption{The absence of the effect of Zeeman splitting on $\kappa_{xy}$. The value of $h_z / \Delta$ is set to $0$, $0.071$, $0.143$, and $0.357$ in
    these four curves.} \label{FigZeeman}
\end{figure}

In Fig.\ 2 in the main text we present the thermal Hall conductivity with no Zeeman term present. The justification for this lies in the fact that
when the result is plotted with the Zeeman term $h_Z < \Delta$ we found no observable change in $\kappa_{xy}$. We illustrate this in Fig.\ \ref{FigZeeman}
where we plot the particular curve of Fig.\ 2, the one onto which other data collapse ($\alpha_D = 7.1$), with the Zeeman term included. The Zeeman
term strength is set to $h_Z / \Delta = 0$, $0.071$, $0.143$, and $0.357$.

